\documentclass[preprint,12pt]{elsarticle}

\usepackage{amssymb}
\usepackage{amsmath}
\usepackage{tikz, lipsum,lmodern}
\usepackage[most]{tcolorbox}
\usepackage{booktabs}
\usepackage{graphicx}
\usepackage{graphicx}%
\usepackage{multirow}%
\usepackage{amsmath,amssymb,amsfonts}%
\usepackage{amsthm}%
\usepackage{mathrsfs}%

\usepackage{xcolor}%
\usepackage{textcomp}%
\usepackage{manyfoot}%
\usepackage{booktabs}%
\usepackage{algorithm}%
\usepackage{algorithmicx}%
\usepackage{algpseudocode}%
\usepackage{listings}%
\usepackage{booktabs}
\usepackage{makecell}
\usepackage{booktabs} 
\usepackage{subcaption}
\usepackage{caption}
\usepackage{fancyhdr}
\usepackage{amsmath} 
\usepackage{array}
\usepackage{graphicx}
 \usepackage{amsmath} 
 \usepackage{booktabs}
\usepackage{graphicx}
\usepackage{booktabs} 
\usepackage{subcaption}
\usepackage{amsmath}
\usepackage{multicol}
\usepackage{mathrsfs}
\usepackage{amssymb}
\usepackage{textcomp}
\usepackage{pdfpages}
\usepackage{wrapfig}
\usepackage{subcaption}
\usepackage{pdflscape}
\usepackage{pdfpages}
\usepackage{float}
\usepackage{afterpage}
\usepackage{tikz, lipsum,lmodern}
\usepackage{calligra,frcursive}
\usepackage{pifont}
\usepackage[most]{tcolorbox}
\usepackage{fancyvrb}
\usepackage{array}
\usepackage{tabularx}
\usepackage{graphicx}  % For \scalebox
\usepackage{booktabs}  % For \midrule, \bottomrule
\usepackage{multirow} 
\usepackage{makecell} 
\usepackage{bbding}
\usepackage{pgfplots}
\usepackage{indentfirst}
\usepackage{xcolor}
\usepackage{rotating}
\usepackage{microtype}
\begin{document}

\begin{frontmatter}

%% Title, authors and addresses

%% use the tnoteref command within \title for footnotes;
%% use the tnotetext command for theassociated footnote;
%% use the fnref command within \author or \affiliation for footnotes;
%% use the fntext command for theassociated footnote;
%% use the corref command within \author for corresponding author footnotes;
%% use the cortext command for theassociated footnote;
%% use the ead command for the email address,
%% and the form \ead[url] for the home page:
\title{{Multicentre Bi-atrial Segmentation from LGE-MRI for Atrial Fibrillation with a 2D and 3D Framework}}

\author[1]{Malitha Gunawardhana\corref{cor1}}
\ead{mgun939@aucklanduni.ac.nz}
\affiliation[1]{organization={Auckland Bioengineering Institute, The University of Auckland},
                addressline={70 Symonds Street},
                city={Auckland},
                postcode={1142}, 
                country={New Zealand}}

\author[1]{Gregory B. Sands}
\ead{g.sands@auckland.ac.nz}

\author[1]{Mark L. Trew }
\ead{m.trew@auckland.ac.nz}

\author[1]{Jichao Zhao}
\ead{j.zhao@auckland.ac.nz}
\cortext[cor1]{Corresponding author: Malitha Gunawardhana}

% \title{A Systematic Analysis for Evaluating Architectural and Dataset Factors in Multi-centre Bi-atrial Segmentation from LGE-MRI for Atrial Fibrillation}

%% use optional labels to link authors explicitly to addresses:
%% \author[label1,label2]{}
%% \affiliation[label1]{organization={},
%%             addressline={},
%%             city={},
%%             postcode={},
%%             state={},
%%             country={}}
%%
%% \affiliation[label2]{organization={},
%%             addressline={},
%%             city={},
%%             postcode={},
%%             state={},
%%             country={}}

% \author[label1]{Malitha Gunawardhana} %% Author name

% %% Author affiliation
% \affiliation[1]{organization={Auckland Bioengineering Institute, The University of Auckland},
%                 addressline={70 Symonds Street},
%                 city={Auckland},
%                 postcode={1142}, 
%                 country={New Zealand}}

%% Abstract
\begin{abstract}
Accurate delineation of bi-atrial structures from late gadolinium enhancement MRI (LGE-MRI) is an important prerequisite for structural analysis and future fibrosis-quantification workflows in atrial fibrillation (AF). However, automated segmentation is
challenging due to thin-walled anatomy, domain shifts across imaging centres, and limited benchmarking of existing methods. This study presents a two-stage segmentation framework and benchmarking platform for evaluating how ROI localisation, encoder design, 2D/3D dimensionality, and ensemble fusion affect bi-atrial wall and cavity segmentation across multicentre LGE-MRI datasets. The framework integrates 3D localisation and fine segmentation using 2D and 3D U-Net variants with ResNeXt encoders and compares them with convolutional, transformer-based, and state-space architectures. { Evaluation across three independent cohorts assessed accuracy and cross-domain transfer without target-domain fine-tuning. Cavity segmentation transferred more consistently across centres than atrial wall segmentation, while wall performance remained sensitive to domain shift, particularly in the Kobe cohort.} By quantifying how 2D, 3D, and ensemble architectures behave across centres and between walls and cavities, this work provides a reproducible benchmark for future methodological development and clinical validation.

\end{abstract}

%%Graphical abstract
\begin{graphicalabstract}
\includegraphics[width=\textwidth]{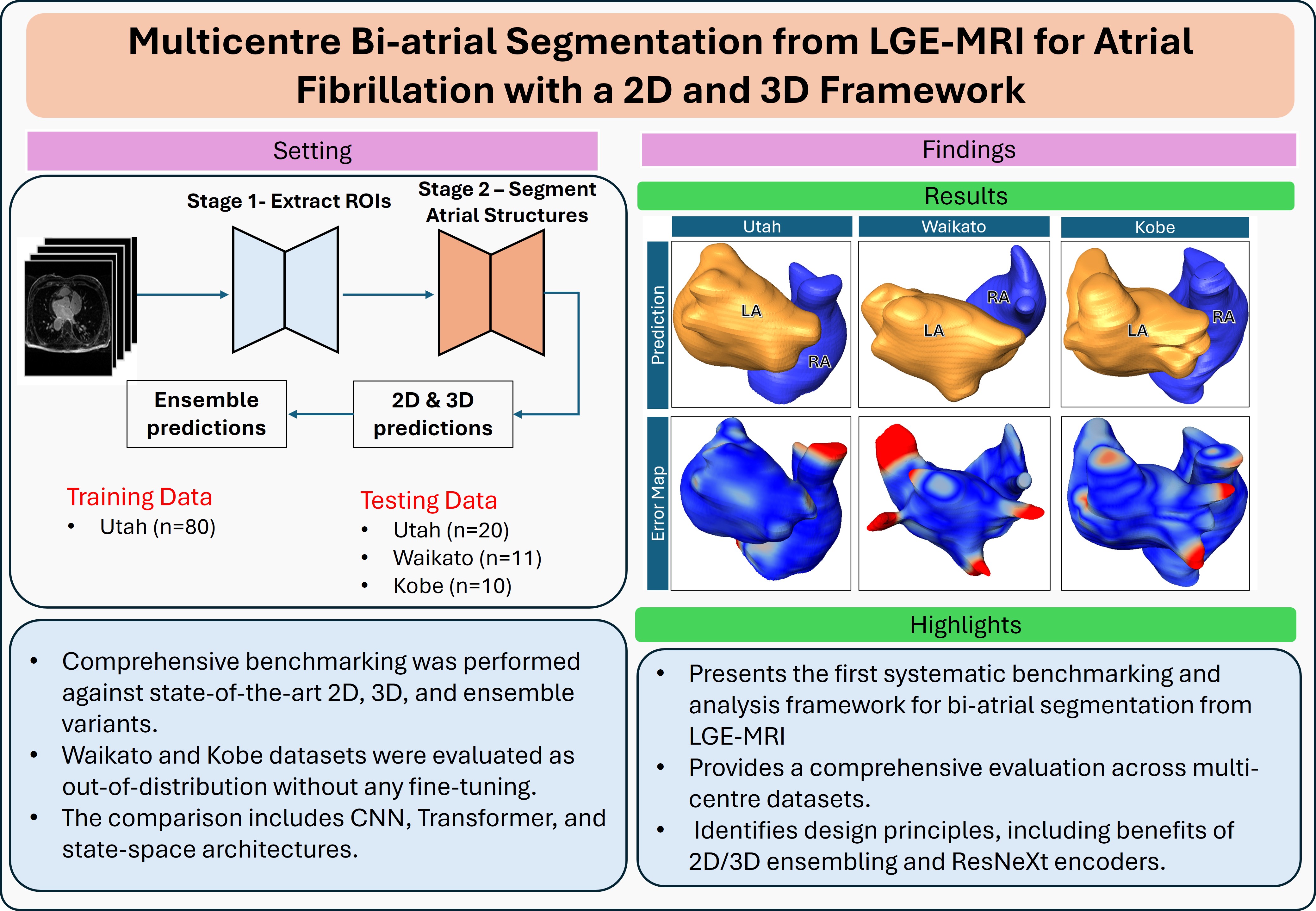}
\end{graphicalabstract}

%%Research highlights
\begin{highlights}
\item  Presents a novel bi-atrial segmentation framework that fuses probability maps from separate 2D and 3D models for LGE-MRI.

\item Evaluates segmentation performance across multiple imaging centres,
analysing robustness under significant domain shift without fine-tuning.
\item Systematically investigates the effects of architectural dimensionality, encoder design, and ensembling on atrial wall and cavity segmentation.
\item {Shows that the hybrid 2D/3D ensemble and ResNeXt encoder provide the most stable overall performance, while atrial wall segmentation remains sensitive to domain shift.}
\item {Provides a reproducible benchmark for future evaluation of structural measurements, fibrosis quantification, and expert-assisted workflows.}
\end{highlights}

%% Keywords
\begin{keyword}
Atrial Fibrillation \sep Segmentation \sep LGE-MRI \sep TASSNet 
%% keywords here, in the form: keyword \sep keyword

%% PACS codes here, in the form: \PACS code \sep code

%% MSC codes here, in the form: \MSC code \sep code
%% or \MSC[2008] code \sep code (2000 is the default)

\end{keyword}

\end{frontmatter}

%% Add \usepackage{lineno} before \begin{document} and uncomment 
%% following line to enable line numbers
%% \linenumbers

%% main text
%%

\section{Introduction}

Atrial fibrillation (AF) is the most common supraventricular arrhythmia worldwide, characterised by rapid, irregular atrial activation that leads to ineffective atrial contraction~\cite{benjamin2019heart,gunawardhana2024integrating}. The condition is associated with increased risks of stroke, heart failure, and mortality, posing a significant burden on global healthcare systems. Despite its prevalence, effective long-term management of AF remains a clinical challenge due to an incomplete understanding of its underlying pathophysiology and the limitations of current therapeutic strategies~\cite{hanna2021assessment, sohns2020atrial}.

Catheter-based pulmonary vein isolation (PVI) remains a cornerstone of interventional AF treatment, particularly for patients with paroxysmal AF. This approach is based on the pivotal role of the pulmonary veins (PVs) in triggering and maintaining AF episodes~\cite{bunch2015pulmonary, simantiris2020cardiac}. However, in cases of persistent AF, PVI alone frequently results in suboptimal outcomes, with high rates of arrhythmia recurrence~\cite{oral2022, kistler2023effect}. These findings underscore the growing recognition that AF is not solely a PV-triggered arrhythmia but is also driven by complex structural and electrical remodelling of the atrial substrate.

One of the key components of this pathological remodelling is atrial fibrosis, non-conductive tissue that can anchor reentrant circuits and perpetuate AF\cite{fan2012cardiac}. Late gadolinium enhancement magnetic resonance imaging (LGE-MRI) enables non-invasive visualisation of fibrotic tissue and has been shown to correlate with disease severity, recurrence risk, and ablation outcomes~\cite{oakes2009detection, hansen2015atrial, higuchi2018spatial, gal2017magnetic, Li2022}. As such, fibrosis imaging has become increasingly important for guiding patient selection and informing individualised ablation strategies.

However, accurate fibrosis quantification depends critically on the \emph{precise segmentation of the atrial cavity and wall}~\cite{feng2025automatic}. Because fibrotic tissue is confined to the thin atrial wall, which borders the blood pool and other tissues of similar intensity, even small segmentation inaccuracies can lead to large errors in fibrosis estimation due to partial volume effects. Manual segmentation is time-consuming, inconsistent, and unsuitable for large-scale or clinical use~\cite{higuchi2018spatial, feng2025automatic}. Consequently, \emph{automated, reproducible segmentation} of the atrial chambers, including both wall and cavity, is essential for reliable fibrosis analysis.

Deep learning, particularly convolutional neural networks (CNNs), has shown promise for automating cardiac segmentation from LGE-MRI~\cite{xiong2018fully, xia2019automatic, xiong2021global, wu2021recent, li2022atrialjsqnet}. CNN-based approaches have achieved strong performance in left atrial (LA) segmentation and fibrosis mapping, improving both efficiency and reproducibility. However, most studies share key limitations: they are trained on small, single-centre datasets, are seldom benchmarked against contemporary architectures, and almost exclusively target the LA~\cite{tao2016fully, yang2018fully, zhuang2023left}.

Despite increasing evidence that AF is a bi-atrial disease involving both the LA and right atrium (RA)~\cite{bax2022effect, kottkamp2013human, Hopman2023}, the RA remains significantly underrepresented in both clinical workflows and research literature. Structural remodelling in the RA has been observed in AF patients, and fibrotic changes in the RA often correlate with those in the LA~\cite{Hopman2023}. Yet, RA analysis is limited by anatomical complexity, thinner wall thickness, and the lack of validated segmentation algorithms tailored to RA morphology~\cite{GunturizBeltran2023, Zhu2024}. More broadly, progress in this field has been hindered by the lack of systematic benchmarking and reproducible frameworks that can evaluate how different architectures, dimensionalities, and dataset characteristics affect segmentation accuracy and transferability. While many studies propose new models, few examine why certain design choices such as dimensionality (2D vs 3D) generalise better across domains or how these behaviours vary between atrial walls and cavities. Consequently, there remains no established reference framework that quantifies model robustness or supports consistent bi-atrial evaluation across centres.

{ 

This gap highlights an urgent need for advanced, generalisable, and fully automated segmentation frameworks capable of accurately delineating both atrial chambers. Reliable segmentation of the LA and RA, including both cavity and wall structures, is essential for comprehensive atrial characterisation, personalised ablation planning, and robust prediction of AF recurrence. Several studies have explored LGE-MRI segmentation (Table~\ref{tab:lge_mri_benchmark_comparison}), although most have focused only on segmenting single structures, such as LA \cite{xiong2021global}, LA scar \cite{zhuang2023left}, or RA cavity~\cite{bai2025benchmark}. MBAS2024~\cite{mbas2024challenge} extended LGE-MRI segmentation towards multiclass bi-atrial analysis. However, its original challenge setting was primarily designed to rank submitted pipelines, and the wall label was treated as a combined bi-atrial wall target rather than as a controlled chamber-specific analysis of LA and RA wall geometry. In such challenge settings, differences in performance may arise from many factors, including preprocessing, augmentation, loss functions, post-processing, ensembling, and implementation details. While challenge rankings are valuable, they do not directly explain which architectural choices drive performance.

Feng et al. ~\cite{feng2025automatic} introduced an automated bi-atrial segmentation pipeline but did not benchmark against current state-of-the-art (SOTA) approaches, raising questions about its relative performance. Moreover, their evaluation on an out-of-distribution dataset involved fine-tuning, which undermines scalability and complicates deployment in diverse clinical environments without site-specific retraining. Critically, their approach employs only three labels for atrial structures (LA and RA cavities plus a unified atrial wall), rather than four (LA/RA cavities and separate LA/RA walls). This simplification discards clinically salient distinctions, as LA and RA walls exhibit divergent functional and pathological profiles in AF. In clinical transfer, separate wall labels enable precise, chamber-specific quantification of atrial wall thickness and fibrosis distribution via targeted post-processing, enhancing spatial fidelity for patient-specific target delineation and reducing propagation errors from shared-wall assumptions at the interatrial septum. Without this granularity, interpretability suffers, and true model capabilities are masked, as aggregated wall predictions inflate Dice scores but dilute biomarker accuracy in multi-centre settings with variable LGE contrast.

\begin{table}[htbp]
\centering
\caption{Comparison with existing LGE-MRI atrial segmentation benchmarks and related studies. 
LA: left atrium; RA: right atrium; LAScar: left atrial scar; MBAS: Multi-class Bi-Atrial Segmentation.}
\label{tab:lge_mri_benchmark_comparison}
\resizebox{\textwidth}{!}{
\begin{tabular}{lll}
\toprule
Study / benchmark & Evaluation type & Label focus \\
\midrule

LA2018~\cite{xiong2021global} 
& Challenge 
& LA cavity\\

LAScarQS2022~\cite{zhuang2023left} 
& Challenge 
& LA cavity, LAScar\\

MBAS2024~\cite{mbas2024challenge} 
& Challenge 
& LA cavity, RA cavity, Bi atrial wall\\

RAS~\cite{bai2025benchmark} 
& Individual benchmark 
& RA cavity 
\\

biAtriaNet~\cite{feng2025automatic} 
& Individual evaluation 
& LA cavity, RA cavity, Bi atrial wall\\

TASSNet (This study) 
& Controlled benchmark 
& LA wall, RA wall, LA cavity, RA cavity 
\\

\bottomrule
\end{tabular}}
\end{table}

}

{
To the best of our knowledge, this study presents TASSNet, a systematic benchmarking and evaluation framework for bi-atrial segmentation from LGE-MRI, unlike prior task-specific methods. The analysis extends beyond the LA by evaluating the LA wall, RA wall, LA cavity, and RA cavity separately. Rather than claiming universal cross-centre robustness, the framework uses controlled comparisons to examine how architectural dimensionality, encoder design, and ensemble fusion influence segmentation accuracy and transferability. Performance is evaluated across three datasets without target-domain fine-tuning, allowing the effects of inter-centre distribution shift to be assessed directly. The results show that cavity segmentation transfers more consistently across centres, whereas atrial wall segmentation remains sensitive to domain shift, particularly in the Kobe cohort. TASSNet therefore provides a reproducible pipeline for separating the effects of architecture, dimensionality, and dataset characteristics while documenting both the strengths and limitations of cross-domain bi-atrial segmentation.
}

Our main contributions are:

\begin{itemize}
\item A systematic benchmarking evaluation for bi-atrial segmentation, integrating both 2D and 3D processing streams with a ResNeXt-based encoder design.
\item A comprehensive cross-domain evaluation across multi-centre datasets, analysing how data heterogeneity and distribution shifts impact model performance.
\item {Empirical characterisation of trade-offs between accuracy, efficiency, and cross-domain stability, including the residual sensitivity of atrial wall segmentation to domain shift.}
\item {A reproducible reference pipeline for future studies of atrial segmentation, harmonisation, domain adaptation, and downstream clinical validation.}

\end{itemize}

{
Rather than focusing only on marginal gains from a single model, this work provides a framework-level analysis of how segmentation architectures behave under consistent experimental conditions. This perspective aims to support more transparent model comparison and provide practical design guidance for future atrial imaging studies. To our knowledge, this is the first LGE-MRI benchmark to evaluate this four-structure task across three independent cohorts without target-domain fine-tuning.}

\section{Methodology}
\begin{figure*}[!t]
\centering
\includegraphics[width=0.8\textwidth]{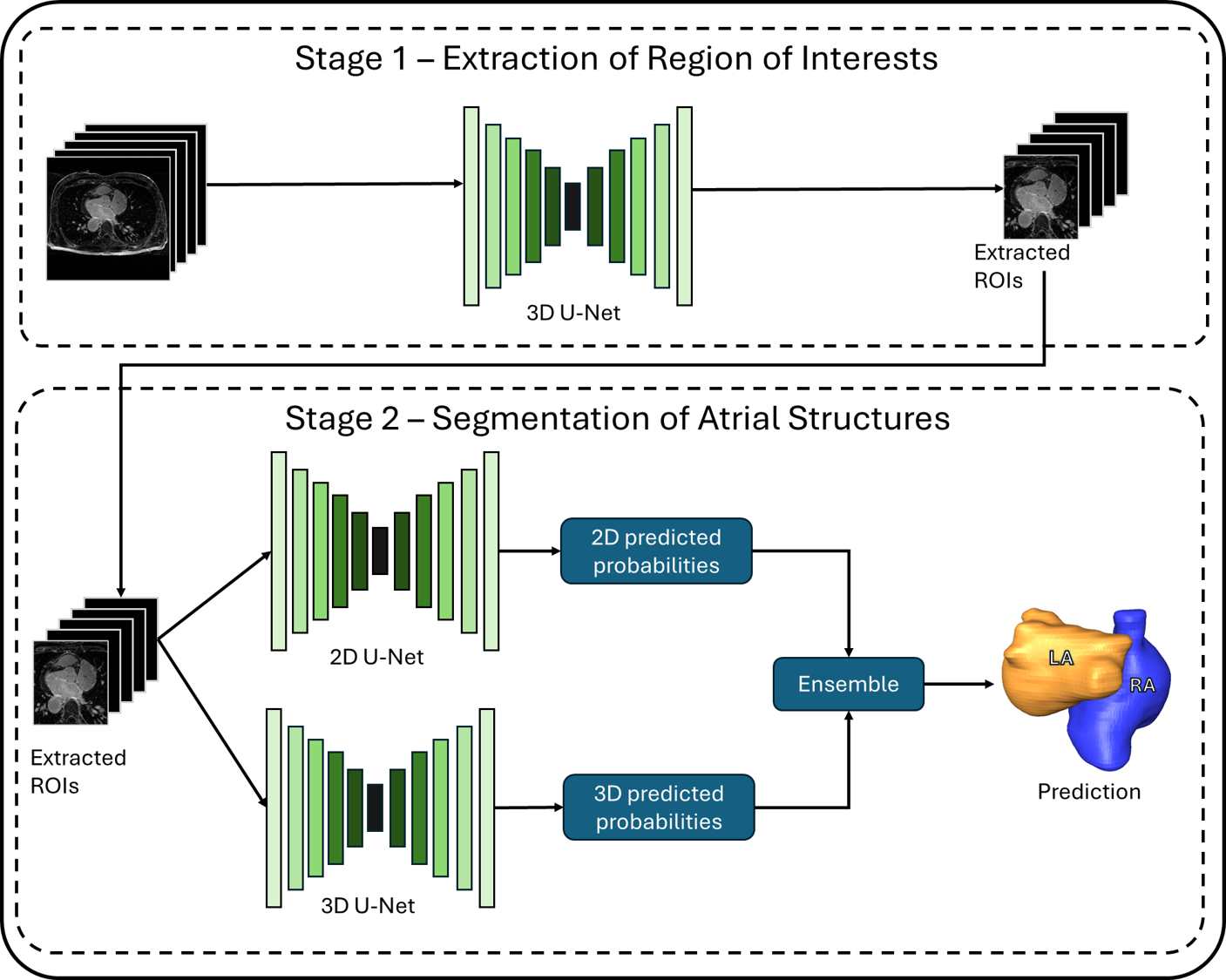}
\caption[Overview of the proposed TASSNet framework]{Overview of the proposed two-stage segmentation framework for atrial structures from 3D LGE-MRI images. In Stage 1, a 3D U-Net is employed to extract regions of interest (ROI) from the input LGE-MRI volume, concentrating on the atrial structures. The output of this stage provides localised ROI, reducing spatial complexity for more precise segmentation in the following stage. In Stage 2, both 2D and 3D U-Net architectures are used to segment the atrial structures within the extracted ROIs finely. The 2D U-Net processes data slice-by-slice, generating 2D probability maps, while the 3D U-Net uses volumetric information to produce 3D probability maps. The final predicted mask, accurately delineating the left and right atrial walls and cavities, is obtained by ensembling the outputs of both networks.}
\label{fig:archi}
\end{figure*}

This study introduces a two-stage deep learning \textbf{benchmarking and segmentation framework} designed to evaluate architectural and dimensional trade-offs in bi-atrial LGE-MRI segmentation. Rather than focusing on a single architecture, the framework acts as an \textit{analytical platform} to assess how dimensionality, encoder structure, and ensemble strategy affect segmentation accuracy and cross-domain behaviour. Figure~\ref{fig:archi} summarises the workflow, which consists of two stages: (1) region-of-interest (ROI) extraction and (2) refined segmentation of atrial structures.

\subsection{Stage 1: Extraction of Regions of Interest (ROIs)}

The first stage leverages a 3D U-Net architecture to perform coarse segmentation of atrial structures. This coarse segmentation serves as a preliminary localisation step designed to identify the general regions of interest within the 3D LGE-MRI volumes that encompass the atria. By focusing only on these regions, the model effectively reduces the computational load and addresses the prevalent issue of class imbalance, which arises from the disproportionate size of the atria relative to the entire MRI volume. 

{
Following coarse segmentation, the centre of mass of the predicted atrial region is computed and used to extract a fixed-size patch of \(256 \times 256 \times 44\) voxels. The crop retains the atrial region while removing most background voxels, thereby reducing class imbalance and the spatial complexity of the fine-segmentation stage. For evaluation, ROI localisation was considered successful when the complete labelled atrial anatomy was contained within the predicted crop without truncation.
}

\subsection{Stage 2: Segmentation of Atrial Structures}

{
Once the ROIs are extracted, the second stage performs refined segmentation of the atrial structures. The two stages were trained independently: the Stage 1 localisation network was trained first and then fixed, and its predicted ROI was used to crop the inputs for the Stage 2 fine-segmentation networks. Two separate U-Nets are employed in Stage 2, one operating on 2D slices and the other on 3D volumes of the extracted ROIs. These complementary configurations capture detailed in-plane information and volumetric spatial context, respectively.
}

Both networks predict the probability that each voxel, or pixel in the 2D configuration, belongs to a specific class. These predictions, represented as probability maps, are then ensembled to produce the final segmentation mask. { The final ensemble prediction was obtained by voxel-wise arithmetic averaging of the 2D and 3D softmax probability maps, followed by an argmax operation.} This ensembling approach combines on the complementary strengths of both 2D and 3D networks. The 2D network excels in capturing fine details on a slice-by-slice basis, while the 3D network is more adept at maintaining spatial coherence across multiple slices, making the combination of their outputs more robust and accurate.

Post-processing is carried out to restore the original dimensions of the predicted segmentation masks. After the initial segmentation in the ROI space, the masks are padded back to the size of the original 3D LGE-MRI input, ensuring that the spatial alignment and anatomical accuracy of the segmentation are preserved.

\subsection{Network Architecture}

The U-Net architecture used in both stages is based on a modified version of the classical U-Net, enhanced with ResNeXt \cite{xie2017aggregated} blocks and instance normalisation (InstanceNorm) to improve convergence during training. The decision to employ ResNeXt blocks instead of the standard block or ResNet \cite{he2016deep} blocks was motivated by the superior ability of ResNeXt to aggregate multiple transformations efficiently, leading to better feature representation and improved performance, especially in high-dimensional medical imaging data. 

While encoders such as ResNet, DenseNet~\cite{huang2017densely},
InceptionNet~\cite{szegedy2015going}, SqueezeNet~\cite{iandola2016squeezenet},
and ConvNeXt~\cite{liu2022convnet} can be used, each comes with limitations. ResNet lacks feature diversity due to its single-path design, while DenseNet is memory-intensive and slower due to dense connections. InceptionNet requires complex, hand-crafted modules that limit scalability. SqueezeNet, though efficient, often sacrifices accuracy, making it less suitable for high-performance tasks. ConvNeXt, although powerful, relies on heavy architectural tuning and is computationally more demanding. ResNeXt encoders introduce cardinality, multiple independent feature subspaces,  that act as implicit domain augmentation, reducing overfitting to scanner-specific texture priors. This design provides a controlled testbed for evaluating the role of feature diversity in cross-domain robustness

The proposed architecture consists of seven stages, with each stage employing convolutional layers with 3x3x3 kernels in the 3D U-Net (and 3x3 kernels in the 2D U-Net). The number of features in the network starts at 32 in the first layer and progressively increases to 64, 128, 256, and eventually 512 in the deeper layers of the network. The final three stages of the encoder maintain a consistent 512 features, ensuring that high-level abstract representations of the input data are captured effectively. This deep, progressive feature extraction allows the network to handle the inherent variability in the shapes and sizes of the atrial structures across different images while also preserving fine-grained details critical for accurate segmentation. The incorporation of ResNeXt blocks further enhances the model’s ability to capture both global and local contextual information, making it well-suited for the complex task of atrial segmentation in medical images.

In the proposed training approach, a cyclical learning rate schedule with exponential decay is used. Let \( R \) represent the total number of epochs and \( Z \) the number of learning rate cycles. The number of epochs per cycle is given by \( T_c = {R}/{Z} \). Within each cycle, the learning rate varies between a maximum value \( lr_r \) and a minimum value \( lr_0 \). The decay rate within a cycle is governed by the scaling factor \( \beta = {M}/{T_c} \), which controls the exponential decrease of the learning rate where \( M \) is an arbitrary number. The learning rate \( lr(i) \) at any epoch \( i \) is defined as follows:

{
\begin{equation}
lr(i)=lr_0 + (lr_r-lr_0)\exp(-\beta t_c),
\end{equation}

where \(t_c = i \bmod T_c\) denotes the epoch index within the current cycle. At the start of each cycle, \(t_c=0\), and therefore \(lr(i)=lr_r\). For subsequent epochs, the learning rate decays exponentially towards the lower bound \(lr_0\). Because the exponential term remains positive for any finite \(t_c\), the schedule approaches \(lr_0\) rather than reaching it exactly. In this study, \(\beta=M/T_c\) with \(M=4\), making the exponential multiplier near the end of each cycle approximately \(\exp(-4)\).

}

\section{Experiments}

\subsection{Implementation Details}

The model was implemented in PyTorch 2.0.1 with a batch size of 4. Training followed a two-stage scheme. The coarse stage performed binary separation of the background and ROI and was trained for 250 epochs. The fine stage performed detailed segmentation and was trained for up to 1000 epochs. A cyclical learning-rate schedule with four cycles (\(Z\)) was used, with the learning rate varying between 0.01 (\(lr_0\)) and 0.1 (\(lr_r\)) and a scaling factor of 4 (\(M\)). Optimisation used AdamW with a weight decay of 0.01 and exponential decay rates of 0.9 and 0.999. DiceFocal loss was selected to address class imbalance~\cite{ma2021loss}. The model used a cardinality value of 8. Evaluation followed five-fold patient-level cross-validation. All experiments were run on a Tesla V100 GPU with 32 GB of memory. The augmentation pipeline included rotation, scaling, Gaussian noise, brightness and contrast adjustment, simulated low resolution, gamma correction with inversion, and mirror flipping. {All baseline models used the same patient-level dataset split, preprocessing procedure, augmentation pipeline, DiceFocal loss, maximum number of training epochs, validation-based model-selection criterion, and evaluation metrics. Input or patch dimensions were adjusted only to accommodate the dimensional requirements of the corresponding 2D or 3D architecture.
}

\subsection{Evaluation Metrics}

The effectiveness of the segmentation model is evaluated using three widely adopted metrics: Dice Similarity Coefficient (DSC), Average Surface Distance (ASD), and the 95th percentile of the Hausdorff Distance (HD95). These metrics collectively provide a comprehensive assessment of the overlap between the predicted and ground truth segmentations, as well as the geometric proximity of their respective surfaces.

\subsection{Datasets}

In this study, three different LGE-MRI datasets were employed in all experiments: (1) Data from the University of Utah~\cite{higuchi2018spatial,oakes2009detection,mcgann2014atrial}, (2) data from the  Waikato Hospital, New Zealand and (3) data from  Kobe University Hospital in Japan. The Waikato and Kobe datasets were evaluated using a model trained solely on the Utah dataset. No additional fine-tuning was performed during testing. A summary of the dataset characteristics is provided in Table~\ref{tab:dataset_summary}, and additional details are presented in the supplementary material. 

Manual reference segmentations were generated for the LA cavity, RA cavity, LA wall, and RA wall using slice-by-slice delineation in Amira. Three observers contributed to the consensus annotation process, and a senior expert subsequently reviewed each final segmentation for anatomical plausibility and consistency. Particular attention was given to the pulmonary veins, atrial appendages, mitral and tricuspid valve planes, interatrial septum, atrial openings, and venous inflow regions. One final consensus segmentation was retained for each scan. Because independent repeated annotations were not available, inter-observer and intra-observer variability could not be quantified and remain limitations of this study.

{ The Utah dataset was split at the patient level into 80 training scans and 20 held-out test scans. The held-out 20 scans were not used for model development and were used for final Utah testing. Within the 80 training scans, five-fold validation was used for model selection and ablation analysis. After model selection, the selected configuration was evaluated on the held-out Utah test set and on the Waikato and Kobe external cohorts. No target-domain fine-tuning was performed for Waikato or Kobe.}

\begin{table}[!h]
\centering
\caption[Summary of the LGE-MRI datasets used in this study]{Summary of the LGE-MRI datasets used in this study, including dataset size, resolution, and source institution.}
\label{tab:dataset_summary}
\resizebox{\columnwidth}{!}{
\begin{tabular}{lcccc}
\toprule
Dataset / Source& Training Images & Testing Images & Z-axis Slices & Final Resolution (mm) \\
\midrule
Utah
& 80 & 20 & 44 & 0.625 × 0.625 × 2.5 \\
Waikato 
& – & 11 & 44 & 0.625 × 0.625 × 2.5 \\
Kobe & – & 10 & 90 & 0.630 × 0.630 × 1.20 \\
\bottomrule
\end{tabular}}
\end{table}

\begin{figure}[ht]
\centering
\begin{subfigure}[t]{0.31\textwidth}
  \centering
  \includegraphics[width=\linewidth]{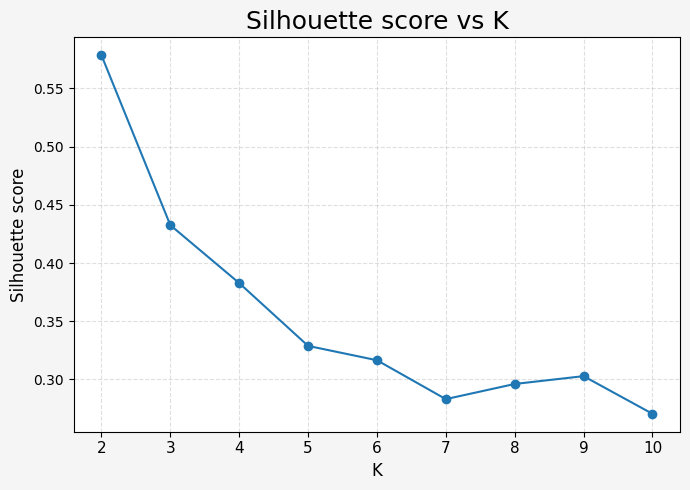}
  \caption{Silhouette score}
  \label{fig:utah_silhouette}
\end{subfigure}\hfill
\begin{subfigure}[t]{0.31\textwidth}
  \centering
  \includegraphics[width=\linewidth]{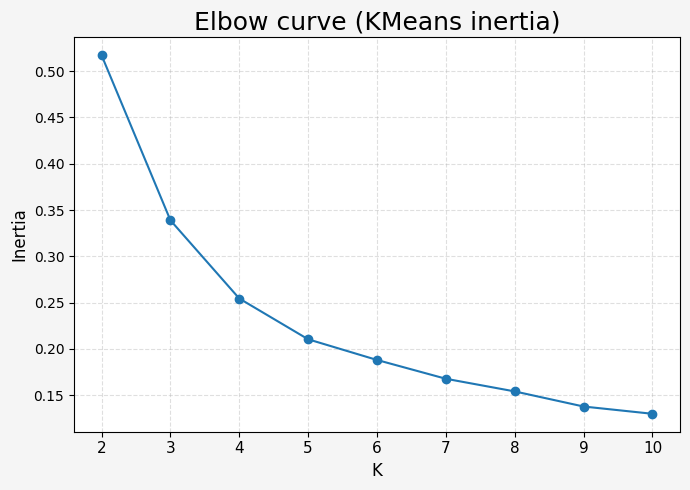}
  \caption{K-means elbow curve}
  \label{fig:utah_elbow}
\end{subfigure}\hfill
\begin{subfigure}[t]{0.31\textwidth}
  \centering
  \includegraphics[width=\linewidth]{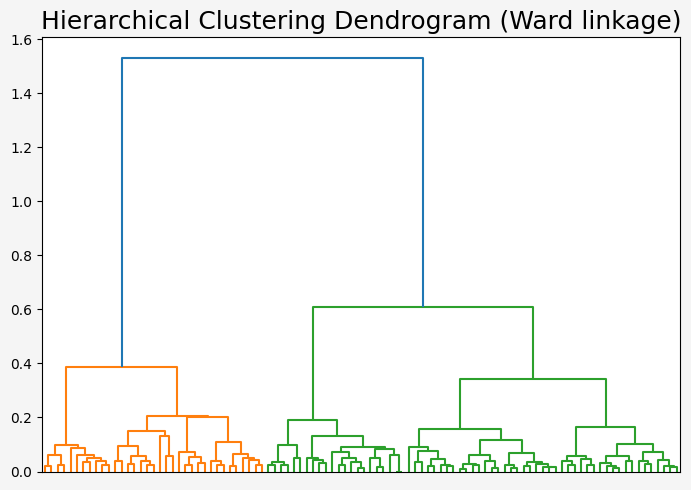}
  \caption{Ward dendrogram}
  \label{fig:utah_ward}
\end{subfigure}

\medskip

\begin{subfigure}[t]{0.31\textwidth}
  \centering
  \includegraphics[width=\linewidth]{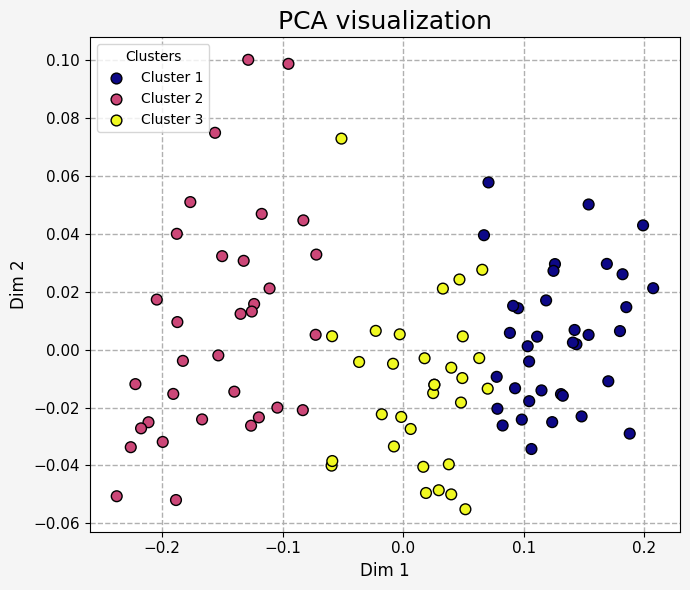}
  \caption{PCA}
  \label{fig:utah_pca}
\end{subfigure}\hfill
\begin{subfigure}[t]{0.31\textwidth}
  \centering
  \includegraphics[width=\linewidth]{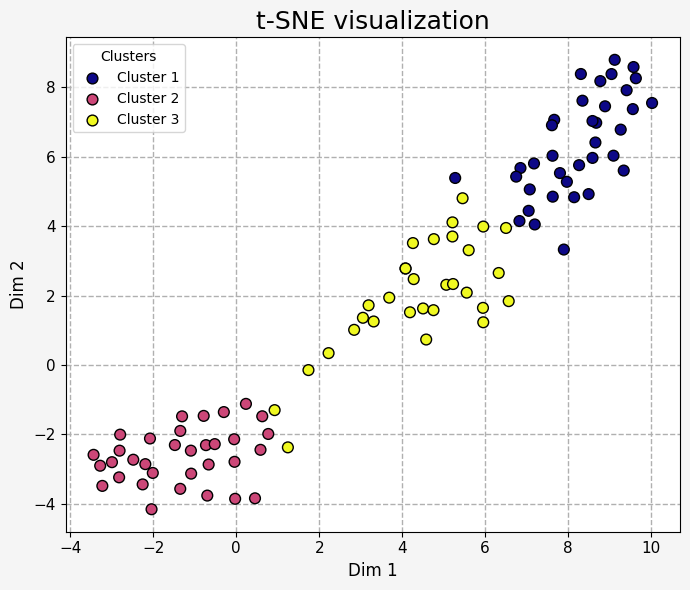}
  \caption{t-SNE}
  \label{fig:utah_tsne}
\end{subfigure}\hfill
\begin{subfigure}[t]{0.31\textwidth}
  \centering
  \includegraphics[width=\linewidth]{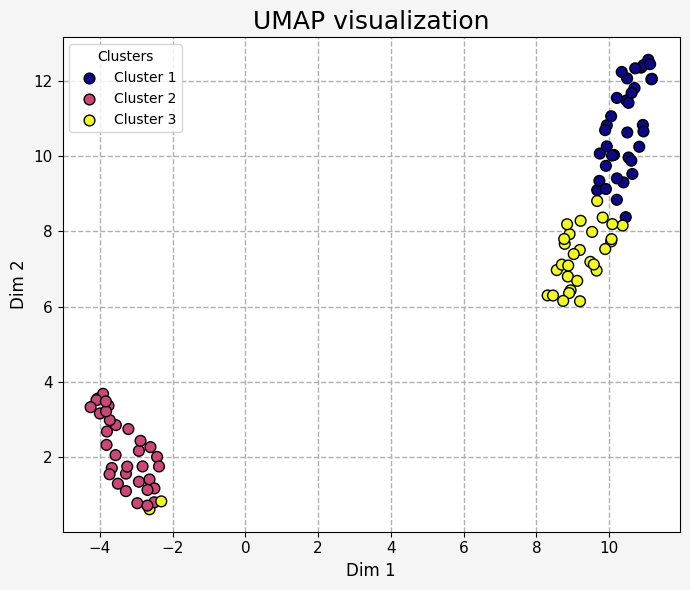}
  \caption{UMAP}
  \label{fig:utah_umap}
\end{subfigure}

\caption{{Utah dataset clustering analysis: (a) silhouette score, (b) K-means elbow curve, (c) Ward hierarchical clustering, and low-dimensional visualisations using (d) PCA, (e) t-SNE, and (f) UMAP.}}
\label{fig:utah_clustering}
\end{figure}

\section{Dataset Analysis}
\label{sec:data_analysis}

To examine the intrinsic structure of the Utah dataset,  a range of clustering validation techniques was used. As shown in the Figure \ref{fig:utah_silhouette}, the silhouette score \cite{rousseeuw1987silhouettes} reached its highest value at $K = 2$ with approximately 0.55, and although it decreased for larger values of $K$, it remained above 0.3 for $K = 3$. This suggests that while two clusters yield the most compact grouping, a three-cluster solution still captures meaningful structure. The elbow method based on K-means inertia showed the steepest decline between $K = 2$ and $K = 3$, after which the curve levelled off, further supporting the choice of three clusters as an optimal balance between compactness and parsimony. Hierarchical clustering using Ward’s method produced a dendrogram that also revealed a natural split into three groups, reinforcing this conclusion.

Nonlinear dimensionality reduction methods provided a visual confirmation of this structure. Both t-SNE \cite{maaten2008visualizing} and UMAP \cite{mcinnes2018umap} generated low-dimensional embeddings in which three distinct clusters were clearly visible. By contrast, PCA \cite{hotelling1933analysis} revealed greater overlap among groups, as expected from a linear projection, but still hinted at the existence of three main clusters. Internal validation indices further corroborated these results. The Davies–Bouldin index \cite{davies2009cluster} was 0.787, which is relatively low and therefore indicative of good cluster separation. In contrast, the Calinski-Harabasz index \cite{calinski1974dendrite} was 196.57, a comparatively high value consistent with compact and distinct clusters. Taken together, these analyses provide convergent evidence that the Utah dataset contains three distinct subclusters. { However, the available metadata is not sufficient for us to determine whether these clusters correspond to acquisition protocol, AF subtype, fibrosis burden, or other clinical characteristics. Therefore, the clustering results are interpreted as evidence of latent dataset heterogeneity rather than clinically defined subgroups.}

\subsection{Dataset Descriptive Statistics and Mutual Information}

\begin{table}[ht]
    \centering
        \caption{Descriptive statistics for Utah, Waikato, and Kobe datasets. Metrics include intensity (mean $\pm$ standard deviation), skewness, kurtosis, label entropy (Ent.), image entropy, and mutual information (MI).}
    \label{tab:dataset_stats}
    \resizebox{\columnwidth}{!}{
    \setlength{\tabcolsep}{3pt}
    \begin{tabular}{lcccccc}
    \toprule
         Dataset & Intensity & Skewness & Kurtosis & Label Ent. & Image Ent. & MI \\
    \midrule
         Utah (Train) & 12.26 $\pm$ 19.73 & 3.33 & 23.91 & 0.1046 & 3.1799 & 0.0324 \\
         Utah (Test) & 8.96 $\pm$ 15.92 & 4.26 & 36.60 & 0.0955 & 2.8530 & 0.0324 \\
    \midrule
         Utah (Combined) & 11.92 $\pm$ 19.40 & 3.43 & 25.19 & 0.1037 & 3.1472 & 0.0325 \\
         Waikato & 22.30 $\pm$ 32.69 & 2.06 & 5.32 & 0.0872 & 3.6704 & 0.0216 \\
         Kobe & 28.44 $\pm$ 41.63 & 1.77 & 3.62 & 0.1319 & 3.0844 & 0.0348 \\
    \bottomrule
    \end{tabular}}

\end{table}

The descriptive statistics in Table~\ref{tab:dataset_stats} highlight distinct differences between datasets. The Utah subsets exhibited high variability, with training intensities averaging 12.26 $\pm$ 19.73 and test intensities averaging 8.96 $\pm$ 15.92. Both were characterised by high skewness (3.33–4.26) and kurtosis (23.9–36.6), suggesting the presence of heavy tails and extreme values. In contrast, the Waikato and Kobe datasets displayed smoother distributions, with higher mean intensities (22.30 and 28.44, respectively) but substantially lower skewness and kurtosis.

Entropy values \cite{shannon1948mathematical} captured dataset complexity. Image entropy was highest for Waikato (3.67), reflecting greater variability, whereas Kobe and Utah both had lower values around 3.1. Label entropy remained low across all datasets (0.08–0.13), indicating relatively imbalanced or homogeneous label distributions. Mutual information values \cite{shannon1948mathematical} were consistently small, ranging from 0.0216 (Waikato) to 0.0348 (Kobe). Although small in magnitude, these differences suggest that Kobe exhibits slightly stronger dependency between features and labels compared with Utah and Waikato. 

{ Additional analyses, including distribution divergence, domain shift assessment, and domain classifier performance, are provided in the Supplementary Material.}

\subsection{Integrated Interpretation}

Together, these analyses provide a comprehensive understanding of dataset characteristics and relationships. The Utah dataset contains three natural clusters confirmed through clustering validation, visualization, and internal indices. Descriptive statistics reveal that Utah is more heterogeneous and has a heavier tail than Kobe or Waikato. Distributional tests highlight Utah versus Kobe as the largest domain shift, while Utah versus Waikato is comparatively closer. The strong performance of the domain classifier confirms that domain differences are meaningful, even when summary metrics suggest only small discrepancies. These findings suggest that while Utah is a heterogeneous and challenging dataset, Waikato may serve as a more suitable external domain for model transfer, whereas Kobe presents a greater shift that requires more careful domain adaptation strategies.

\section{Results}

The proposed TASSNet was evaluated against several SOTA segmentation
models and their respective variants. These models include CNN-based architectures such as the original nnU-Net framework \cite{isensee2021nnu,gunawardhana2024good}, nnU-Net with ResNet encoders (ResEncM), LightM-UNet \cite{liao2024lightm}, and SegResNet \cite{myronenko20193d}. Transformer-based architectures were also considered, including Swin UNETR \cite{hatamizadeh2021swin} and two variations of SAMed with patch sizes of 256 and 512 \cite{zhang2023customized}. Additionally, state-space model-based architectures, such as U-MambaBot and U-MambaEnc \cite{ma2024u}, were included. Two variations of TASSNet were also evaluated using ResNet and ResNeXt encoders.

In nnU-Net, the “3D” configuration refers to a 3D U-Net operating at full image resolution, while “3D low resolution” operates on downsampled images. Additionally, a 3D U-Net cascade is employed, where the first network processes downsampled images and a second network refines its segmentation outputs at full resolution. In the U-MambaBot architecture, a Mamba-based bottleneck layer is incorporated between the encoder and decoder, whereas U-MambaEnc implements its entire encoder using Mamba layers. Beyond these models, the evaluation also considered additional variants for comparative analysis. Additional results are available in the supplementary material. 

{
The Stage 1 predicted ROI contained the complete atrial anatomy in all 41 held-out test scans, comprising 20 Utah, 11 Waikato, and 10 Kobe scans. This corresponds to an ROI localisation success rate of 100\%, with no test case classified as a crop failure.
}

\subsection{Quantitative Performance}

Table~\ref{tab:utah_dataset} presents the performance on the Utah dataset test set, while Table~\ref{tab:waikato_dataset} summarises the results for the Waikato dataset. Table~\ref{tab:kobe_datasets} reports the performance on the Kobe datasets. The best results for each evaluation metric are highlighted in \textcolor{red}{red}. For detailed statistical significance analyses, please refer to the supplementary materials, where results are provided for all three datasets. Additional analyses comparing 2D, 3D, and ensemble models are also included in the supplementary material for each dataset.

\begin{table*}[!h]
\centering
\caption[Performance of different methods on the Utah dataset]{Performance of different methods on the Utah dataset for Right Atria (RA) wall, Left Atria (LA) wall, RA cavity, and LA cavity. DSC: Dice Similarity Coefficient, ASD: Average Surface Distance, HD95: 95th percentile of Hausdorff Distance. The best values are \textcolor{red}{highlighted}. 3D low:3D low resolution, 3D cas:3D cascade model, Ens: Ensemble. }
\label{tab:utah_dataset}
\setlength{\tabcolsep}{3pt}
\resizebox{1\textwidth}{!}{
\begin{tabular}{p{2.5cm}|l|p{1.3cm}|p{1.3cm}|p{1.3cm}|p{1.3cm}|p{1.3cm}|p{1.3cm}|p{1.3cm}|p{1.3cm}|p{1.3cm}|p{1.3cm}|p{1.3cm}|p{1.3cm}}

 \toprule 
\multirow{2}{*}{Model}&\multirow{2}{*}{Config}& \multicolumn{3}{c|}{RA Wall}& \multicolumn{3}{c|}{LA wall}& \multicolumn{3}{c|}{RA cavity}& \multicolumn{3}{c}{LA cavity}\\ && DSC& ASD& HD95& DSC& ASD& HD95& DSC& ASD& HD95& DSC& ASD&HD95\\ 
\midrule

\multirow{5}{*}{nnUNet}
&2D & 0.738 & 0.548 & 2.094 & 0.594 & 0.771 & 3.125 & 0.915 & 0.769 & 3.031 & 0.918 & 0.773 & 3.098 \\ 
&3D & 0.735 & 0.557 & 2.069 & 0.609 & {0.739} & \textcolor{red}{2.778} & 0.913 & 0.771 & 2.786 & 0.920 & 0.758 & 2.866 \\ 
&3D low& 0.723 & 0.622 & 2.408 & 0.574 & 0.973 & 4.231 & 0.908 & 0.841 & 3.094 & 0.917 & 0.821 & 3.216 \\ 
&3D cas.& 0.735 & 0.556 & 1.981 & 0.609 & {0.739} & 2.797 & 0.914 & 0.779 & 2.751 & 0.920 & 0.758 & 2.916 \\ 
&Ens. & 0.740 & 0.541& 2.040 & 0.610 & 0.751 & 2.856 & 0.915 & 0.762 & 2.817 & 0.921 & 0.750 & 2.949 \\

\midrule
\multirow{5}{*}{\makecell[l]{nnUNet \\ with ResEnc}} &2D 
& 0.734& 0.551& 2.076& 0.592& 0.741& 2.857& 0.915& 0.741& 2.860& 0.919& 0.756& 2.869\\ 
&3D & 0.737& 0.531& \textcolor{red}{1.920} & 0.606&0.744& 2.819& 0.913& 0.765& 2.812& 0.921& 0.741& 2.776\\ 
&3D low& 0.712& 0.656& 2.603& 0.555& 1.016& 4.378& 0.906& 0.862& 3.163& 0.916& 0.817& 3.050\\ 
&3D cas.& 0.736& 0.548& 1.952& 0.600& 0.760& 2.881& 0.913& 0.760& 2.765& 0.920& 0.754& 2.808\\ 
 &Ens. & 0.740& 0.538& 1.933&0.605& 0.748& 2.815& 0.915& 0.751& 2.739& 0.922& 0.739& 2.835\\

\midrule
\multirow{3}{*}{UMambaBot}& 2D& 0.730& 0.547& 2.039& 0.598& 0.733& 2.913& 0.912& 0.789& 3.045& 0.918& 0.765&2.905
\\
 & 3D & 0.735& 0.548& 2.087& 0.612& 0.725& 2.814& 0.912& 0.765& 2.738& 0.921& 0.731&2.733
\\
 & Ens.& 0.743& \textcolor{red}{0.514}& 1.939& 0.613& 0.703& 2.771& 0.918& 0.724& 2.667& 0.922& 0.705&2.709\\

 \midrule
 \multirow{3}{*}{UMambaEnc}& 2D
& 0.730& 0.561& 2.137& 0.601& 0.709& 2.799& 0.908& 0.869& 3.482& 0.920& 0.733&2.859
\\
 & 3D 
& 0.737& 0.556& 2.107& 0.617& 0.729& 2.743& 0.913& 0.768& 2.802& 0.921& 0.742&2.777
\\
 & Ens.& 0.746& 0.522& 1.945& 0.618& \textcolor{red}{0.696}& 2.771& 0.916& 0.762& 2.897& 0.922& \textcolor{red}{0.685}&\textcolor{red}{2.620}\\ 

 \midrule
 \multirow{3}{*}{\makecell[l]{LightM \\ Unet}}& 2D
& 0.681& 0.879& 3.881& 0.556& 0.965& 3.841& 0.884& 1.142& 4.565& 0.902& 1.005&4.049
\\
 & 3D 
& 0.709& 0.680& 2.606& 0.586& 0.858& 3.329& 0.896& 0.947& 3.688& 0.912& 0.891&3.557
\\
 & Ens.& 0.715& 0.721& 3.132& 0.582& 0.887& 3.531& 0.905& 0.895& 3.447& 0.913& 0.886&3.653\\ 

 \midrule

\multirow{3}{*}{\makecell[l]{Swin \\ UNETR }} 
 & 2D& 0.694& 0.715& 2.782& 0.554& 0.899& 3.568& 0.900& 0.982& 3.765& 0.903& 0.931&3.613
\\
 & 3D & 0.716& 0.640& 2.444& 0.604& 0.830& 3.218& 0.903& 0.879& 3.270& 0.915& 0.845&3.240
\\
 & Ens.& 0.722& 0.623& 2.419& 0.587& 0.804& 3.165& 0.913& 0.785& 3.094& 0.912& 0.824&3.277\\

\midrule
\multirow{3}{*}{SegResNet}
& 2D& 0.717& 0.683& 3.133& 0.572& 0.877& 3.538& 0.901& 0.917& 3.786& 0.909& 0.946&3.501\\
& 3D & 0.728& 0.621& 2.323& 0.611& 0.756& 2.864& 0.908& 0.812& 2.988& 0.922& 0.749&2.877\\
& Ens.& 0.738& 0.609& 2.500& 0.596& 0.813& 3.286& 0.916& 0.759& 2.863& 0.920& 0.768&3.021\\
\midrule
SAMed-B & 2d(256) & 0.709& 0.706& 2.675& 0.565& 0.919& 3.763& 0.894& 0.870& 3.816& 0.913& 0.813&3.039
\\
\midrule
 SAMed-H& 2d(512) & 0.718& 0.650& 2.451& 0.553& 0.939& 3.925& 0.901& 0.908& 3.703& 0.913& 0.842&3.307\\

 \midrule
\multirow{3}{*}{\makecell[l]{TASSNet \\ with ResNet}}&2D & 0.737& 0.555& 2.082& 0.606& 0.772& 3.034& 0.914& 0.766& 2.832& 0.920& 0.761& 2.949
\\
&3D & 0.738& 0.547& 2.038& 0.612& 0.745& 2.854& 0.915& 0.758& 2.750& 0.921& 0.744& 2.901
\\
&Ens. & 0.743& 0.545& 2.075& 0.616& 0.746& 2.923& 0.919& 0.719& 2.675& 0.922& 0.728& 2.815
\\

\midrule

  \multirow{3}{*}{\makecell[l]{TASSNet\\with ResNeXt}}
    & 2D   & 0.742 & 0.562 & 2.225 & 0.608 & 0.757 & 3.073 & 0.915 & 0.756 & 3.017 & 0.918 & 0.772 & 2.914 \\

    & 3D   & 0.739 & 0.641 & 2.350 & 0.617 & 0.838 & 3.421 & 0.913 & 0.815 & 3.093 & 0.919 & 0.812 & 3.232 \\

    & Ens.\ & \textcolor{red}{0.753} & 0.564 & 2.159 & \textcolor{red}{0.620} & 0.739 & 3.080 & \textcolor{red}{0.920} & \textcolor{red}{0.713} & \textcolor{red}{2.655} & \textcolor{red}{0.923} & 0.702 & 2.684 \\

  \bottomrule
\end{tabular}}
\end{table*}

\subsubsection{Utah Dataset}

On the Utah dataset, which represents the in-domain scenario, both TASSNet configurations (ResNet and ResNeXt) achieved comparable performance to strong baselines such as nnU-Net and U-Mamba variants. However, the ResNeXt-based TASSNet ensemble yielded the most balanced performance across all four anatomical substructures, particularly for thin-walled regions where local continuity is critical. Compared to transformer and Mamba-based models, which showed slightly lower wall Dice and higher surface errors, TASSNet’s consistent accuracy highlights the effectiveness of its hybrid 2D–3D representation and moderate encoder cardinality. This suggests that architectural simplicity, when paired with representational diversity, can achieve robust in-domain segmentation without resorting to heavily parameterised architectures.

\subsubsection{Waikato Dataset}

Similar to the observations on the Utah dataset, the results for the Waikato dataset, summarised in Table~\ref{tab:waikato_dataset}, indicate that TASSNet with ResNeXt achieved the highest Dice scores for segmenting the RA wall, RA cavity, and LA cavity. In contrast, TASSNet with ResNet demonstrated the best performance for the LA wall. For the ASD and HD95 metrics, alternative models such as nnU-Net with ResNet encoders, UMambaEnc, and SAMed-H achieved superior results.

\begin{table*}[!h]
\centering
\setlength{\tabcolsep}{3pt}
\caption[Performance of different methods on the Waikato dataset]{Performance of different methods on the Waikato dataset for Right Atria (RA) wall, Left Atria (LA) wall, RA cavity, and LA cavity. DSC: Dice Similarity Coefficient, ASD: Average Surface Distance, HD95: 95th percentile of Hausdorff Distance. The best values are \textcolor{red}{highlighted}. 3D low:3D low resolution, 3D cas:3D cascade model, Ens: Ensemble. }
\label{tab:waikato_dataset}
\resizebox{1\textwidth}{!}{
\begin{tabular}{p{2.5cm}|l|p{1.3cm}|p{1.3cm}|p{1.3cm}|p{1.3cm}|p{1.3cm}|p{1.3cm}|p{1.3cm}|p{1.3cm}|p{1.3cm}|p{1.3cm}|p{1.3cm}|p{1.3cm}}

 \toprule 
\multirow{2}{*}{Model}&\multirow{2}{*}{Config}& \multicolumn{3}{c|}{RA Wall}& \multicolumn{3}{c|}{LA wall}& \multicolumn{3}{c|}{RA cavity}& \multicolumn{3}{c}{LA cavity}\\ && DSC& ASD& HD95& DSC& ASD& HD95& DSC& ASD& HD95& DSC& ASD&HD95\\ 
\midrule

\multirow{5}{*}{nnUNet}
 &2D & 0.608& 1.724& 6.738& 0.470& 2.098& 8.897& 0.785& 1.847& 6.491& 0.816& 2.542& 11.737
\\ 
 &3D & 0.667& 1.009& 4.543& 0.560& 1.198& 5.283& 0.846& 1.292& 4.522& 0.867& 2.046& 10.008
\\ 
 &3D low& 0.637& 1.100& 4.699& 0.507& 1.891& 9.127& 0.857& 1.282& 4.206& 0.847& 2.662& 13.585
\\ 
 &3D cas.& 0.669& 0.961& 4.095& 0.547& 1.356& 6.293& 0.862& 1.184& 4.030& 0.864& 2.192& 11.017
\\ 
 &Ens. & 0.676& 0.837& 3.217& 0.559& 1.209& 5.397& 0.863& 1.177& 3.934& 0.866& 2.075& 10.295\\ 

\midrule
\multirow{5}{*}{\makecell[l]{nnUNet \\ with ResEnc}} &2D 
& 0.592& 1.858& 7.113& 0.484& 1.695& 7.379& 0.781& 1.784& 5.861& 0.835& 2.257& 10.199
\\ 
 &3D 
& 0.678& 0.820& 3.208& 0.566&1.136& 4.920& 0.863& 1.141& 3.754& 0.869& \textcolor{red}{1.952}& 9.467
\\ 
 &3D low
& 0.626& 1.071& 4.322& 0.487& 1.775& 7.917& 0.858& 1.268& 3.999& 0.856& 2.432& 12.229
\\ 
 &3D cas.
& 0.672& \textcolor{red}{0.810}& \textcolor{red}{3.033}& 0.555& 1.184& 5.230& 0.863& 1.164& 3.782& 0.869& 2.007& 9.975
\\ 
 &Ens. & 0.667& 0.886& 3.501&0.544& 1.312& 6.081& 0.866& 1.144& \textcolor{red}{3.742}& 0.869& 2.073& 10.435
\\

\midrule
\multirow{3}{*}{UMambaBot}
 & 2D& 0.621& 1.346& 5.184& 0.500& 1.539& 6.885& 0.815& 1.523& 4.966& 0.832& 2.201&9.878
\\
 & 3D & 0.676& 0.879& 3.558& 0.559& 1.176& 5.242& 0.860& 1.199& 4.073& 0.868& 2.044&10.203
\\
 & Ens. & 0.652& 1.076& 4.579& 0.526& 1.476& 6.831& 0.844& 1.290& 4.377& 0.851& 2.103&9.958
\\

 \midrule
 \multirow{3}{*}{UMambaEnc}& 2D
& 0.644& 1.007& 4.175& 0.496& 1.570& 6.872& 0.837& 1.407& 4.935& 0.840& 2.065&8.998
\\
 & 3D 
& 0.668& 0.943& 3.886& 0.557& \textcolor{red}{1.148}& \textcolor{red}{5.031}& 0.854& 1.276& 4.443& 0.862& 2.114&10.060
\\
 & Ens.& 0.658& 0.966& 3.921& 0.524& 1.399& 6.282& 0.851& 1.268& 4.394& 0.857& 2.018&9.418\\ 
 
 \midrule
\multirow{3}{*}{\makecell[l]{LightM\\ Unet}}& 2D
& 0.491& 3.301& 12.148& 0.466& 1.906& 8.047& 0.749& 2.016& 6.569& 0.816& 2.512&11.755
\\
 & 3D 
& 0.449& 10.595& 28.390& 0.503& 2.090& 9.089& 0.614& 4.841& 15.063& 0.823& 3.371&17.733
\\
 & Ens.& 0.481& 3.747& 13.579& 0.484& 2.049& 8.694& 0.717& 2.552& 9.082& 0.830& 2.778&14.222\\ 
 \midrule

\multirow{3}{*}{\makecell[l]{Swin \\ UNETR}}
 & 2D& 0.584& 1.637& 6.586& 0.460& 1.543& 6.757& 0.799& 1.696& 5.989& 0.834& 2.077&9.017
\\
 & 3D & 0.617& 1.155& 4.995& 0.539& 1.386& 6.166& 0.844& 1.348& 4.553& 0.848& 2.482&12.762
\\
 & Ens.& 0.603& 1.222& 5.412& 0.496& 1.516& 6.692& 0.830& 1.412& 4.933& 0.851& 2.177&10.665\\

\midrule
\multirow{3}{*}{SegResNet}
& 2D& 0.518& 2.437& 8.778& 0.462& 1.921& 7.686& 0.727& 2.061& 6.793& 0.814& 2.536&11.539
\\
 & 3D & 0.525& 5.179& 16.534& 0.533& 2.704& 9.784& 0.680& 4.832& 14.424& 0.838& 3.237&15.434
\\
 & Ens.& 0.507& 3.386& 11.586& 0.497& 2.485& 10.181& 0.710& 2.245& 7.119& 0.834& 2.793&13.775\\

\midrule
SAMed-B & 2d(256) & 0.585& 2.000& 8.708& 0.456& 3.421& 13.453& 0.779& 1.901& 6.810& 0.817& 2.630&12.803
\\
\midrule
 SAMed-H & 2d(512) & 0.636& 1.223& 5.215& 0.488& 1.589& 6.901& 0.833& 1.497& 5.806& 0.859& 1.866&\textcolor{red}{8.676}\\

 \midrule
\multirow{3}{*}{\makecell[l]{TASSNet \\ with ResNet}}&2D & 0.667& 0.927& 3.764& \textcolor{red}{0.570}& 1.402& 6.675& 0.864& 1.195& 3.882& 0.862& 2.319& 11.695\\ 
&3D & 0.652& 1.124& 4.992& 0.548& 1.631& 7.735& 0.861& 1.215& 4.175& 0.857& 2.490& 12.639\\ 
&Ens. & 0.671& 0.990& 4.251&{0.561}& 1.429& 6.831& 0.863& 1.180& 3.986& 0.868& 2.155& 11.080
\\ \midrule
 \multirow{3}{*}{\makecell[l]{TASSNet \\ with ResNeXt}}
&2D & \textcolor{red}{0.681}& 0.841& 3.351& {0.566}& 1.169& 5.176& 0.862& 1.171& 4.057& 0.869& 2.030& 10.298\\  

&3D & 0.677& 0.856& 3.379& 0.553& 1.279& 5.650& 0.861& 1.201& 4.144& 0.867& 2.143& 11.116\\  

&Ens. & \textcolor{red}{0.681}& 0.839& 3.270& {0.561}& 1.198& 5.334& \textcolor{red}{0.867}& \textcolor{red}{1.138}& 4.030& \textcolor{red}{0.870}& 1.992& 9.880\\ 
 \bottomrule
\end{tabular}}
\end{table*}

Evaluation on the Waikato dataset, unseen during training, demonstrates the framework’s ability to test cross-centre consistency. Here, all models exhibit an expected performance drop relative to Utah, reflecting differences in acquisition and demographics. Yet, TASSNet with ResNeXt maintains the highest mean Dice score across atrial structures, confirming that hybrid ensembling mitigates degradation under moderate distribution shifts. The larger variance observed in wall Dice scores indicates that wall regions are more sensitive to centre-specific contrast variations than cavities, aligning with clinical expectations. Interestingly, 2D configurations sometimes outperformed 3D ones on this smaller dataset, implying that models with reduced contextual dependency may be less biased by scanner anisotropy.

\subsubsection{Kobe Dataset}

\begin{table*}[!h]
\centering
\setlength{\tabcolsep}{3pt}
\caption[Performance of different methods on the Kobe dataset]{Performance of different methods on the Kobe dataset for Right Atria (RA) wall,
Left Atria (LA) wall, RA cavity, and LA cavity. DSC: Dice Similarity Coefficient, ASD: Average
Surface Distance, HD95: 95th percentile of Hausdorff Distance. The best values are highlighted.
3D low:3D low resolution, 3D cas:3D cascade model, Ens: Ensemble. }
\label{tab:kobe_datasets}
\resizebox{1\textwidth}{!}{
\begin{tabular}{p{2.5cm}|l|p{1.3cm}|p{1.3cm}|p{1.3cm}|p{1.3cm}|p{1.3cm}|p{1.3cm}|p{1.3cm}|p{1.3cm}|p{1.3cm}|p{1.3cm}|p{1.3cm}|p{1.3cm}}

 \toprule 
\multirow{2}{*}{Model}&\multirow{2}{*}{Config}& \multicolumn{3}{c|}{RA Wall}& \multicolumn{3}{c|}{LA wall}& \multicolumn{3}{c|}{RA cavity}& \multicolumn{3}{c}{LA cavity}\\ && DSC& ASD& HD95& DSC& ASD& HD95& DSC& ASD& HD95& DSC& ASD&HD95\\ 
\midrule

\multirow{5}{*}{nnUNet}
 &2D & 0.444& 4.710& 17.779& 0.431& 3.681& 12.139& 0.778& 4.079& 19.364& 0.852& 2.384& 8.295
\\ 
 &3D & 0.502& 2.822& 13.625& 0.440& 4.077& 12.964& 0.800& 2.580& 9.067& 0.837& 2.596& 9.258
\\ 
 &3D low& 0.454& 5.735& 21.432& 0.472& 2.971& 13.037& 0.703& 4.660& 16.492& 0.861& 2.570& 10.017
\\ 
 &3D cas.& 0.481& 3.328& 14.127& 0.457& 3.599& 12.093& 0.790& 3.066& 10.379& 0.795& 5.219& 27.413
\\ 
 &Ens. & 0.464& 5.285& 18.483& 0.436& 4.414& 15.975& 0.753& 4.161& 12.495& 0.789& 4.671& 20.766
\\ 

\midrule
\multirow{5}{*}{\makecell[l]{nnUNet \\ with ResEnc}} &2D 
& 0.497& 3.125& 14.566& 0.491& 1.670& 8.118& 0.816& 2.693& 10.231& 0.834& 4.457& 22.258
\\ 
 &3D 
& 0.455& 4.824& 18.546& 0.446&4.658& 13.370& 0.729& 4.330& 13.550& 0.805& 4.387& 21.800
\\ 
 &3D low
& 0.415& 7.568& 23.378& 0.423& 4.500& 18.877& 0.675& 5.290& 18.907& 0.749& 6.437& 29.860
\\ 
 &3D cas.
& 0.450& 6.321& 19.702& 0.519& \textcolor{red}{1.408}& \textcolor{red}{6.447}& 0.735& 7.714& 16.791& 0.871& 4.004& 23.185
\\ 
 &Ens. & 0.458& 5.924& 20.526&0.472& 2.752& 12.840& 0.741& 4.400& 13.172& 0.764& 7.895& 26.559
\\

\midrule
\multirow{3}{*}{UMambaBot}
 & 2D& 0.502& 2.672& 13.212& 0.464& 2.469& 10.017& 0.826& 2.437& 9.379& 0.860& 2.302&8.019
\\
 & 3D & 0.465& 4.003& 18.384& 0.455& 4.472& 12.538& 0.754& 3.436& 12.311& 0.816& 3.735&13.728
\\
 & Ens.& 0.474& 4.296& 17.428& 0.454& 3.026& 11.818& 0.800& 2.996& 10.805& 0.863& 2.433&9.412
\\

 \midrule
 \multirow{3}{*}{UMambaEnc}& 2D
& 0.486& \textcolor{red}{2.389}& \textcolor{red}{11.368}& 0.479& 1.898& 8.507& 0.826& 2.453& 9.148& 0.878& 2.002&7.589
\\
 & 3D 
& 0.397& 11.890& 32.620& 0.409& 7.675& 18.169& 0.655& 8.354& 21.511& 0.704& 6.238&24.385
\\
 & Ens.& 0.463& 4.294& 17.076& 0.435& 3.408& 13.139& 0.781& 3.116& 11.260& 0.832& 2.728&11.280
\\ 
 
 \midrule
\multirow{3}{*}{\makecell[l]{LightM\\ Unet}}& 2D
& 0.438& 3.749& 16.512& 0.403& 3.593& 13.337& 0.785& 3.236& 11.385& 0.825& 3.206&11.529
\\
 & 3D 
& 0.406& 7.918& 27.261& 0.375& 7.083& 22.930& 0.735& 4.019& 13.490& 0.786& 6.375&20.719
\\
 & Ens.& 0.427& 4.516& 18.347& 0.409& 4.353& 16.140& 0.781& 3.185& 11.421& 0.842& 3.502&16.758
\\ 
 \midrule
\multirow{3}{*}{\makecell[l]{Swin \\ UNETR}}
 & 2D& 0.436& 3.255& 14.456& 0.422& 4.544& 13.884& 0.761& 3.155& 11.434& 0.809& 3.582&11.584
\\
 & 3D & 0.396& 10.720& 29.125& 0.437& 2.511& 11.727& 0.699& 4.890& 17.564& 0.798& 5.999&30.795
\\
 & Ens.& 0.409& 9.403& 25.068& 0.435& 2.939& 12.077& 0.699& 5.490& 16.821& 0.853& 3.851&19.712
\\

\midrule
\multirow{3}{*}{SegResNet}
& 2D& 0.455& 4.243& 17.005& 0.435& 3.683& 11.845& 0.784& 2.960& 9.978& 0.834& 3.322&12.328
\\
 & 3D & 0.277& 16.751& 38.950& 0.362& 6.097& 23.123& 0.526& 7.605& 21.048& 0.776& 5.006&15.163
\\
 & Ens.& 0.312& 14.430& 36.066& 0.447& 3.167& 13.146& 0.609& 9.331& 24.127& 0.858& 2.955&14.528
\\

\midrule
SAMed-B & 2d(256) & 0.491& 3.132& 15.801& 0.439& 3.136& 12.306& 0.814& 2.508& 9.579& 0.860& 2.306&7.673
\\
\midrule
 SAMed-H & 2d(512) & 
0.509& 2.883& 15.005& 0.449& 1.977& 9.294& 0.816& 2.547& 9.274& 0.887& 2.019&7.789
\\

 \midrule
\multirow{3}{*}{\makecell[l]{TASSNet \\ with ResNet}}&2D & 0.497& 3.775& 17.617& 0.477& 2.453& 12.097& 0.808& 2.687& 10.237& 0.847& 3.908& 19.406
\\ 
&3D & 
0.484& 4.363& 18.929& 0.484& 2.453& 10.823& 0.788& 2.875& 10.388& 0.875& 2.502& 10.671
\\ 
&Ens. & 0.496& 3.678& 17.244&0.483& 2.713& 13.228& 0.804& 2.770& 10.664& 0.832& 4.758& 24.562
\\ \midrule

 \multirow{3}{*}{\makecell[l]{TASSNet \\ with ResNeXt}}
&2D & 
0.487& 3.733& 18.184& 
0.491& 2.498& 
10.450& 0.817& 2.595& 9.564& 0.867& 2.225& 8.162
\\    
&3D & 0.488& 3.469& 16.664& 0.487& 2.399& 11.065&0.806& 2.741& 9.763& 0.888& 2.078& 7.963
\\    
&Ens. & 
\textcolor{red}{0.511}& 2.960& 14.726& \textcolor{red}{0.521}& 2.326& 11.253& \textcolor{red}{0.832}& \textcolor{red}{2.350}& \textcolor{red}{8.046}& \textcolor{red}{0.902}& \textcolor{red}{1.640}& \textcolor{red}{6.039}
\\ 
 \bottomrule

\end{tabular}}

\end{table*}

{
The Kobe dataset presents the most distinct acquisition domain, with the largest MMD and KS divergence from Utah. All methods showed greater performance degradation in this cohort, particularly for atrial wall segmentation. TASSNet with ResNeXt achieved the highest Dice values for the four evaluated structures, including an LA cavity Dice of 0.902, but wall Dice remained substantially lower than cavity Dice. The stronger relative performance of several 2D configurations suggests that slice-wise representations may be less affected by differences in through-plane resolution. These observations support the use of feature-diverse and hybrid configurations as candidate strategies for cross-centre evaluation, while also showing that they do not eliminate wall-segmentation sensitivity to domain shift.
}

\begin{figure}[!h]
    \centering
    \includegraphics[width=\linewidth]{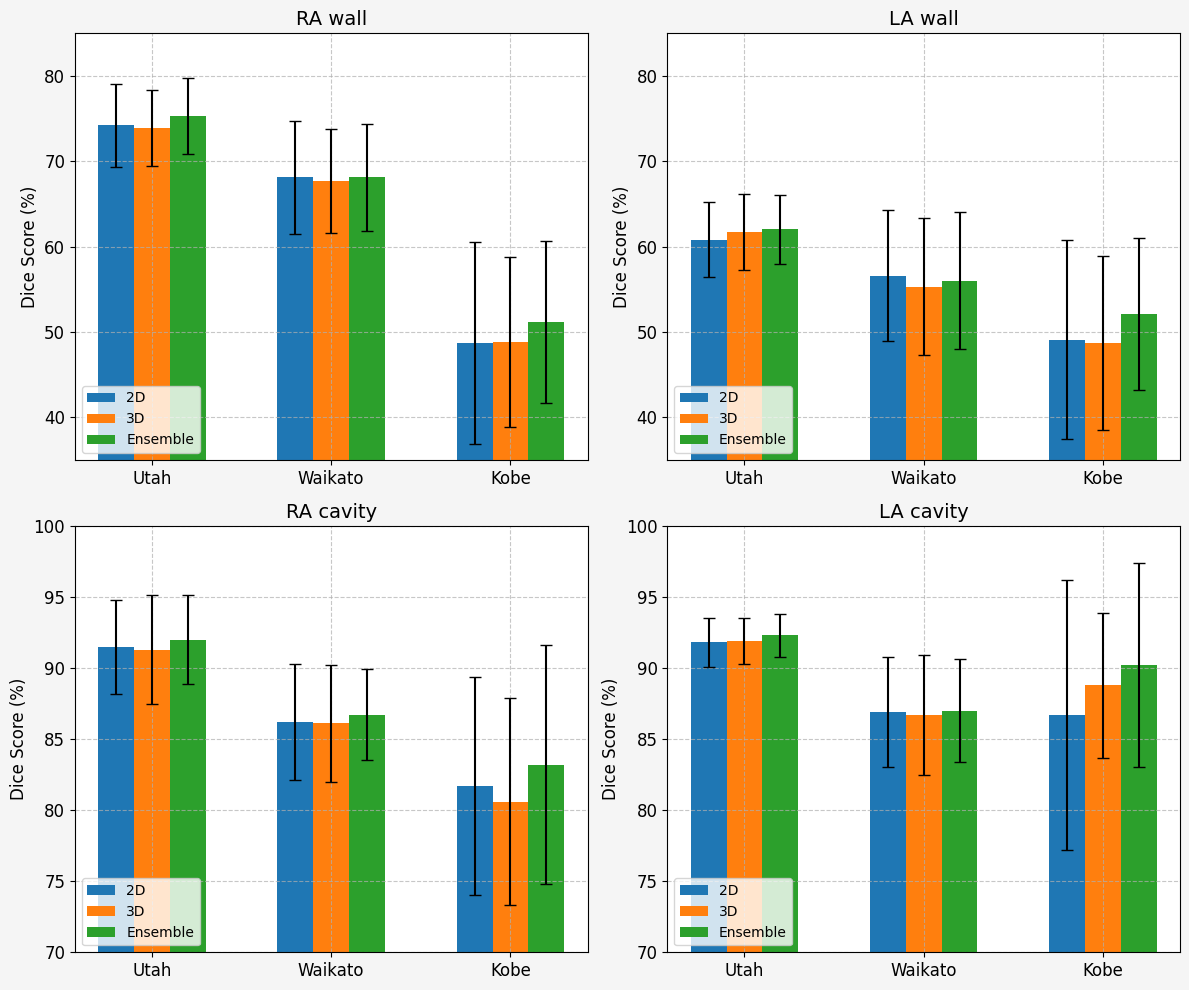}
    \caption[Overall performance of TASSNet with ResNeXt encoders across all three datasets]{Overall performance of TASSNet with ResNeXt encoders across all three datasets, evaluated using 2D, 3D, and ensemble configurations for the left atrial (LA) and right atrial (RA) cavity and wall.}
    \label{fig:overall}
\end{figure}

Figure~\ref{fig:overall} provides a detailed comparison of the segmentation performance of TASSNet with ResNeXt encoders across all three datasets: Utah, Waikato, and Kobe. The evaluation is stratified by RA wall, LA wall, RA cavity, and LA cavity. For each structure, segmentation accuracy is assessed using three model configurations: 2D, 3D, and an ensemble of both. Dice scores are reported as percentages, and standard deviations are represented by error bars to reflect the consistency of model predictions across individual test cases.

{
The results reveal consistent trends across datasets. The RA and LA cavities achieved higher Dice scores and generally lower variability than the atrial walls. This performance gap was especially evident in the Waikato and Kobe cohorts. For the TASSNet ResNeXt ensemble, RA wall Dice decreased from 75.3\% on Utah to 51.1\% on Kobe, while LA wall Dice decreased from 62.0\% to 52.1\%. The corresponding cavity reductions were smaller, from 92.0\% to 83.2\% for the RA cavity and from 92.3\% to 90.2\% for the LA cavity. Patient-level bootstrap analysis also showed greater uncertainty for wall segmentation. On Waikato, the ensemble Dice was 0.681 [0.639, 0.723] for the RA wall and 0.561 [0.507, 0.615] for the LA wall, compared with 0.867 [0.845, 0.889] and 0.870 [0.846, 0.894] for the RA and LA cavities. On Kobe, the wall Dice values were 0.511 [0.442, 0.580] and 0.521 [0.457, 0.585], whereas the cavity Dice values were 0.832 [0.772, 0.892] and 0.902 [0.851, 0.953]. Full bootstrap 95\% confidence intervals for Dice, ASD, and HD95 are reported in the Supplementary Material.
}

\subsubsection{Inference time analysis}

Table~\ref{tab:run_tim} summarises the average inference times for both 2D and 3D models across the three datasets. For the Utah and Waikato datasets, inference times remain relatively low and consistent between the two configurations, with the 3D models taking slightly longer than their 2D counterparts. In contrast, the Kobe dataset shows notably higher inference times for both configurations, with the 3D model requiring substantially more time. This increase reflects the greater computational demand associated with the dataset's larger input volume.

\begin{table}[!h]
    \centering
        \caption[Average inference time (in seconds)]{Average inference time (in seconds) for 2D and 3D models across different datasets, with standard deviation.}
    \label{tab:run_tim}
    \begin{tabular}{lcc}
    \toprule
         Dataset & \multicolumn{2}{c}{Inference Time (s)} \\
         \midrule
         & 2D & 3D \\
         Utah & 2.38 $\pm$ 0.68& 4.91 $\pm$ 1.33\\
         Waikato & 2.51 $\pm$ 0.67& 4.71 $\pm$ 1.56\\
         Kobe & 5.47 $\pm$ 0.91& 9.81 $\pm$ 2.74\\
         \bottomrule
    \end{tabular}

\end{table}

\subsubsection{Slice by Slice Performance}

\begin{figure}[!h]
    \centering
    \includegraphics[width=\linewidth]{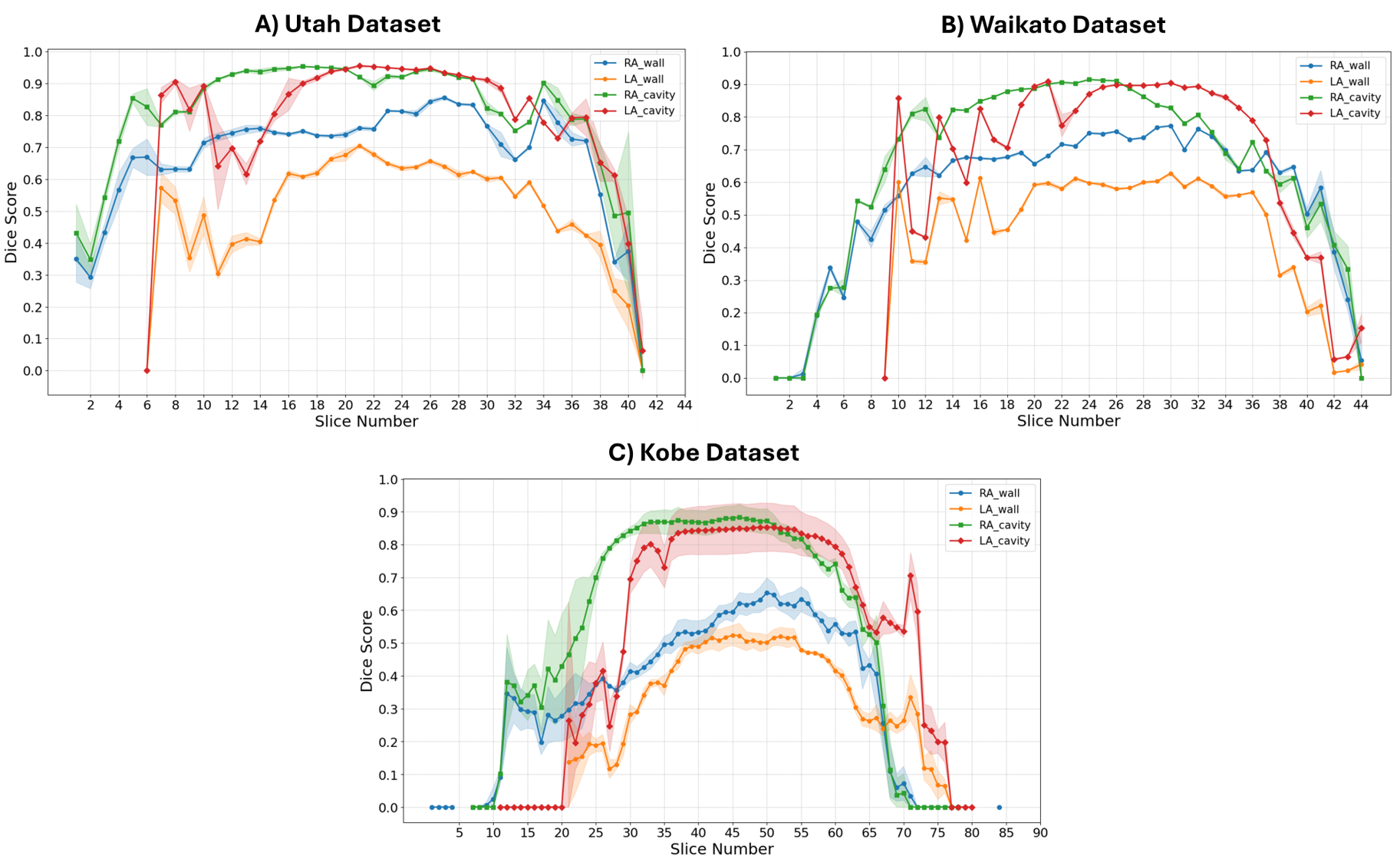}
    \caption[Slice-by-slice evaluation of the predicted performance using dice score]{Slice-by-slice evaluation of the predicted performance using dice score. Values are averaged across the 2D, 3D and Ensemble models from the TASSNet with ResNeXt. LA- Left Atria, RA- Right Atria.}
    \label{fig:slice_by_slice}
\end{figure}

Figure~\ref{fig:slice_by_slice} presents a slice-wise Dice score analysis of segmentation performance across the RA and LA walls and cavities, averaged over the 2D, 3D, and ensemble configurations of TASSNet with a ResNeXt backbone.

In the Utah dataset (Figure~\ref{fig:slice_by_slice}A), the LA is absent in the early slices (0–6), and the models correctly predict no structure in these regions, resulting in no reported Dice scores for the LA structures. The RA, however, is present from the initial slices, with progressively improving segmentation accuracy. From approximately slices 8 to 32, all anatomical structures become more clearly defined, resulting in a marked rise in Dice scores. Peak performance is observed in this central region, with Dice scores exceeding 0.9 for both the RA and LA cavities. In contrast, segmentation accuracy remains consistently lower for the atrial walls, particularly the LA wall, likely due to its thin and irregular morphology, which presents greater challenges for accurate segmentation. Beyond slice 36, a decline in Dice scores is observed across all structures, corresponding with reduced anatomical visibility in the inferior portions of the volume.

A similar pattern is observed in the Waikato dataset (Figure~\ref{fig:slice_by_slice}B). In the earliest slices, RA Dice scores were near zero, but segmentation performance improved in the mid-slices before declining again in the final slices. The overall Dice score trend is comparable to that observed in the Utah dataset.

In the Kobe dataset (Figure~\ref{fig:slice_by_slice}C), segmentation performance in the initial slices was poor, with Dice scores close to zero and inconsistent predictions, alternating between correctly predicting no structure and incorrectly predicting segmentation where no structure was present. In the final slices, the RA wall was absent in the ground truth, yet a single slice showed a false positive prediction by the model. Similar to the Utah and Waikato datasets, segmentation performance improved considerably in the mid-slices.

Slice-wise analysis further clarifies where segmentation performance degrades anatomically. High Dice scores in mid-atrial regions coincide with maximal wall visibility and stable contrast, while apical and basal slices suffer from lower signal and more ambiguous boundaries. These patterns mirror known LGE-MRI artefacts and validate that observed errors arise from intrinsic anatomical and imaging factors rather than model instability.

\subsection{Qualitative Performance}

\begin{figure}[!htp]
\centering
\includegraphics[width=0.8\columnwidth]{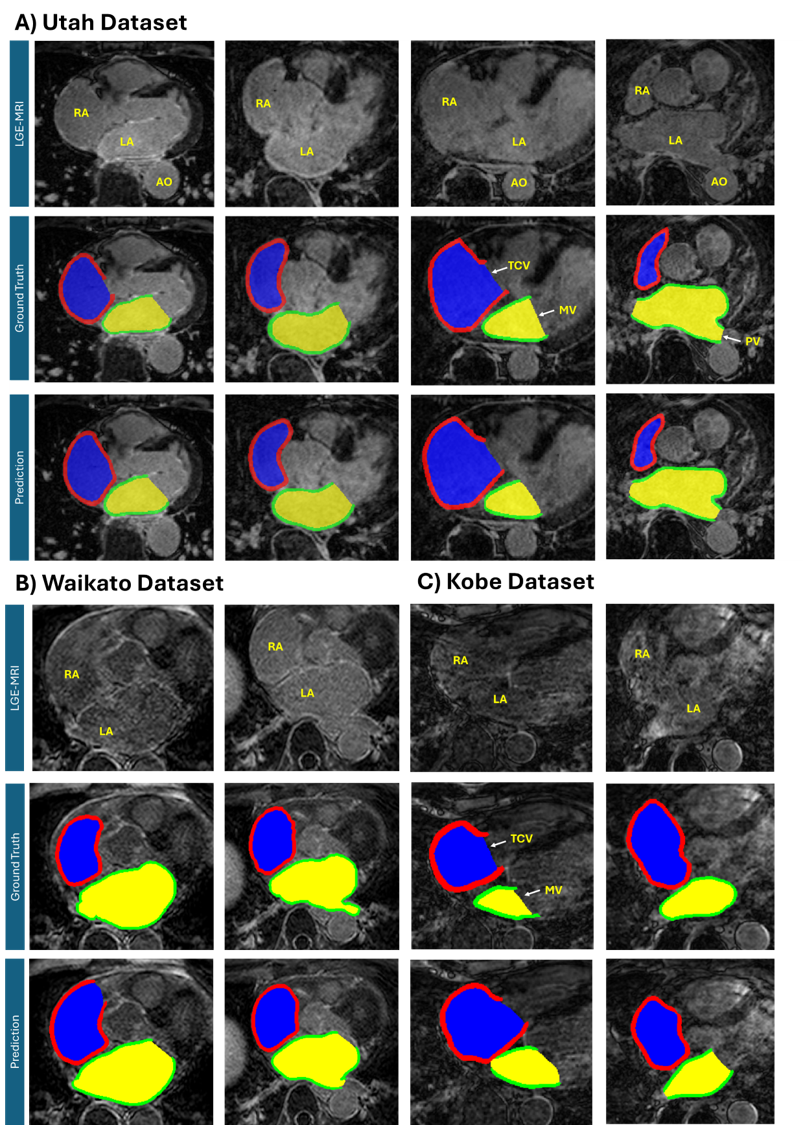}
\caption[Comparison of left atria (LA) and right atria (RA) wall and cavity segmentation ]{Comparison of Left Atria (LA) and Right Atria (RA) Wall and Cavity Segmentation from MRI
Segmentation results are shown for three datasets: A) Utah, B) Waikato, and C) Kobe. Each column represents a different MRI scan. LA cavity (yellow), RA cavity (blue), LA wall (green), and RA wall (red). Key anatomical structures are labelled: AO – aorta; MV/TCV – mitral/tricuspid valve; PV – pulmonary vein.}
\label{fig:qualitative}
\end{figure}

Figure~\ref{fig:qualitative} illustrates qualitative segmentation results for the LA and RA walls and cavities across representative LGE-MRI scans for the Utah dataset. Each column corresponds to a different subject, and each row presents the raw MRI image, manual ground truth, and model prediction, respectively. The segmented anatomical structures are colour-coded as follows: LA cavity (yellow), RA cavity (blue), LA wall (green), and RA wall (red). The figure illustrates the model’s ability to accurately delineate complex atrial anatomy, allowing for a visual assessment of the segmentation quality in comparison to expert annotations. Key anatomical landmarks, such as the aorta (AO), mitral and tricuspid valves (MV/TCV), and pulmonary veins (PV), are also labelled to assist in interpretation.

\begin{figure}[!h]
    \centering
    \includegraphics[width=\columnwidth]{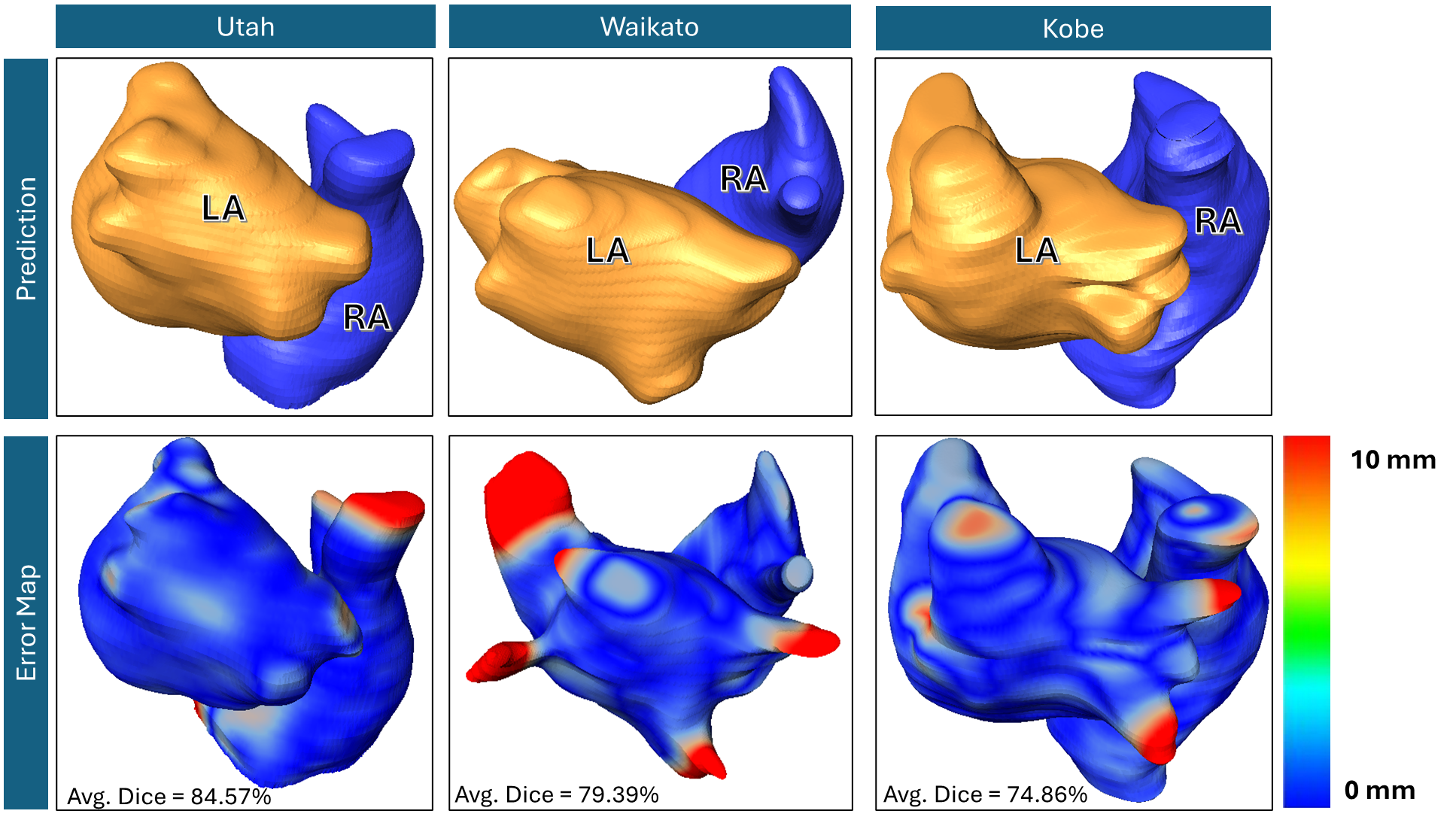}
    \caption[Comparison of 3D views]{Three-dimensional visualisation of the predicted left and right atria (LA and RA) for one scan from each dataset. The first row shows the model predictions, while the second row presents the corresponding error maps overlaid on the ground truth segmentations. Average Dice scores are shown in the bottom left corner of the error maps.}
    \label{fig:3d_views}
\end{figure}

In Figure~\ref{fig:3d_views}, three-dimensional reconstructions are presented for representative scans from each of the three datasets. The first row displays the model predictions, while the second row shows the corresponding error maps overlaid on the ground truth segmentations. The average Dice scores, computed across the four anatomical structures, are reported in the bottom left corner of each error map. As the error maps are superimposed on the ground truth, a separate visualisation of the ground truth segmentation is not provided. The figure also highlights that the model encounters challenges in accurately segmenting the atrial opening regions.

\subsubsection{2D, 3D and Ensemble Performance}

\begin{figure}[!h]
    \centering
    \includegraphics[width=\columnwidth]{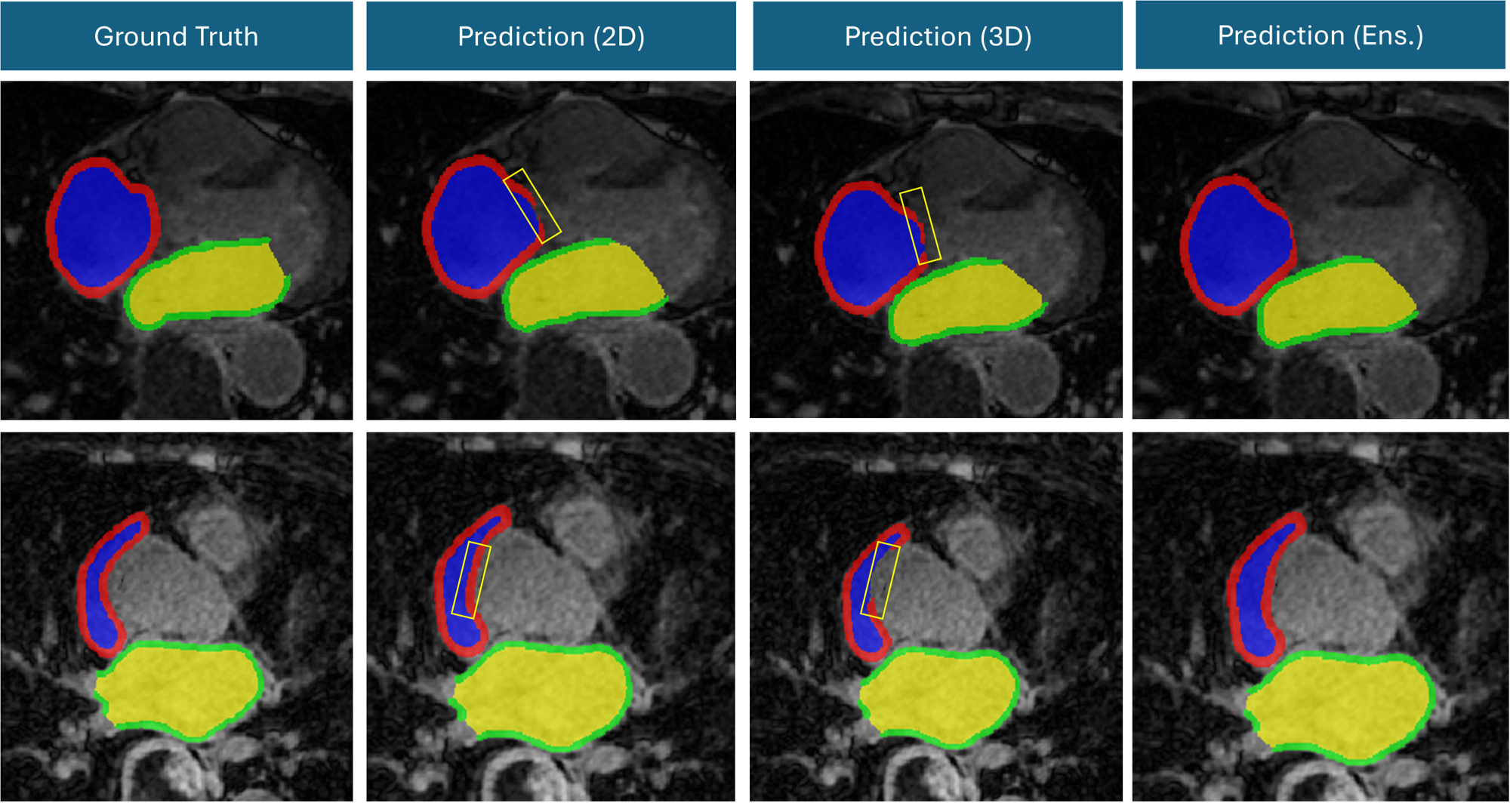}
\caption[Comparison of segmentation predictions from 2D, 3D, and ensemble models]{Comparison of segmentation predictions from 2D (first column), 3D (second column), and ensemble model (third column). Each row corresponds to a different LGE-MRI scan. The segmented anatomical structures are colour-coded as follows: left atrial (LA) cavity in yellow, right atrial (RA) cavity in blue, LA wall in green, and RA wall in red. The yellow bounding boxes highlight regions where noticeable differences are observed between 2D and 3D model predictions, illustrating the variability in anatomical boundary delineation across configurations.}
    \label{fig:2d_3d_change}
\end{figure}

Figure~\ref{fig:2d_3d_change} compares segmentation predictions generated by the 2D, 3D, and ensemble variants of TASSNet equipped with a ResNeXt backbone. Each row showcases a different LGE-MRI slice, with corresponding predictions shown side by side. In certain cases, which highlighted with yellow bounding box, the 3D model outperforms the 2D model (e.g., first row), while in others, the 2D model achieves better results (e.g., second row). The ensemble model effectively integrates the strengths of both approaches, leading to improved overall performance.

\subsubsection{Failure Cases}

\begin{figure}[!h]
\centering
\includegraphics[width=\columnwidth]{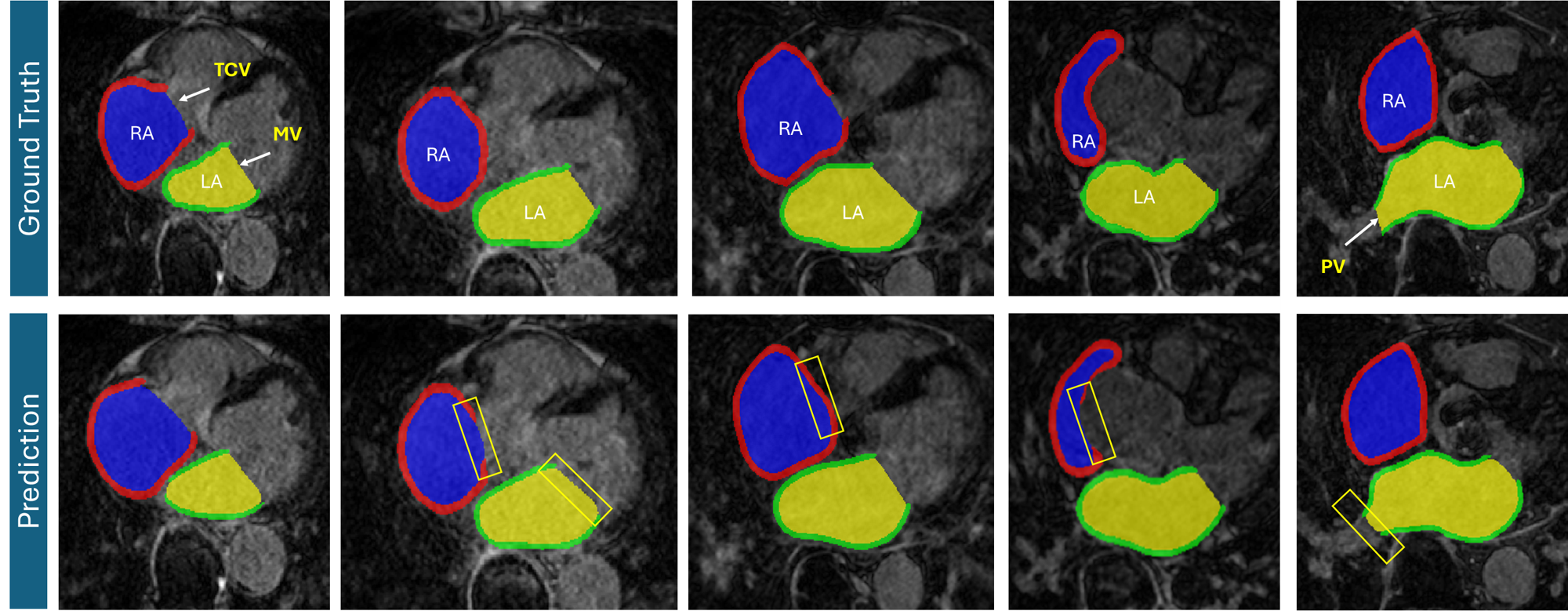}
\caption[Visualisation of failure cases]{Visualisation of failure cases. Each column corresponds to a different LGE-MRI scan. The segmented LA cavity is shown in yellow, RA cavity in blue, LA wall in green, and RA wall in red. Yellow bounding boxes highlight regions of segmentation failure, particularly in atrial wall regions, in comparison to the ground truth.}
\label{fig:failure_cases}
\end{figure}

{
Failure cases are illustrated in Figure~\ref{fig:failure_cases}, showing examples where the model produced suboptimal segmentations in anatomically complex or low-contrast regions. Overall, cavity segmentation remained relatively accurate across most slices, whereas atrial wall delineation was more challenging. Errors were mainly localised around anatomical openings, including the tricuspid valve, mitral valve, and pulmonary veins, where weak contrast, intensity inhomogeneity, and ambiguous anatomical boundaries reduce wall visibility. Common error patterns included leakage or missing boundaries around the pulmonary veins, boundary shifts near the mitral and tricuspid valve planes, and local discontinuities around atrial openings. Thin LA and RA wall regions were particularly sensitive to these errors because partial volume effects and the small number of wall voxels mean that even minor missed segments can affect Dice and surface-distance metrics. Larger boundary errors were more evident in the Kobe scans, consistent with the stronger domain shift and acquisition differences observed for this cohort.
}

{\subsection{Downstream Quantitative Analyses}}
\subsubsection{Atrial Volume Measurement}

In addition to anatomical segmentation, TASSNet is applied to the quantitative evaluation of atrial volume estimation. Accurate volume measurements are essential in the diagnosis, monitoring, and treatment planning of AF, where precise assessment of atrial remodelling and tissue characteristics supports informed clinical decision-making and effective patient stratification~\cite{raniga2024left, matei2022early, tzeis2019atrial}. Traditionally, such measurements are performed manually or semi-automatically from imaging data, introducing inter-observer variability and limiting reproducibility. By leveraging the precise anatomical segmentation produced by TASSNet, atrial volumes can be computed in a consistent, reproducible, and scalable fashion. The voxel-wise delineation of atrial cavities provided by the segmentation masks enables geometry-based calculations without the need for manual input.

\begin{table}[!h]
    \centering
    \caption[Quantitative assessment of atrial volume measurements (in mL) for the Utah, Waikato, and Kobe datasets]{Quantitative assessment of atrial volume measurements (in mL) for the Utah, Waikato, and Kobe datasets, covering both left atrium (LA) and right atrium (RA). Metrics include Mean Absolute Error (MAE), Root Mean Squared Error (RMSE), Mean Absolute Percentage Error (MAPE), Pearson correlation coefficient ($\rho$), Regression Coefficient (RC), and bias.}
    \label{tab:volume}
    \resizebox{0.6\columnwidth}{!}{
    \begin{tabular}{lcccccc}
    \toprule
         Metric & \multicolumn{2}{c}{Utah} & \multicolumn{2}{c}{Waikato} & \multicolumn{2}{c}{Kobe} \\
         \cmidrule{2-7}
         & LA & RA & LA & RA & LA & RA \\
         \midrule
         MAE & 7.098 & 5.809 & 9.075 & 8.948 & 9.161 & 15.852 \\
         RMSE & 8.232 & 7.496 & 13.605 & 10.923 & 14.634 & 19.384 \\
         MAPE (\%) & 8.340 & 7.885 & 8.578 & 12.727 & 9.869 & 15.713 \\
         $\rho$ & 0.947 & 0.981 & 0.984 & 0.849 & 0.913 & 0.818 \\
         RC & 0.768 & 1.056 & 0.717 & 0.756 & 0.960 & 0.666 \\
         Bias & -1.227 & 0.596 & -6.004 & 3.509 & -3.555 & -13.341 \\
         \bottomrule
    \end{tabular}}
\end{table}

For volume estimation, all values are in millilitres (mL). A lower  Mean Absolute Error (MAE), Root Mean Squared Error (RMSE), Mean Absolute Percentage Error (MAPE), and bias indicate better predictive accuracy, while higher values of  Pearson correlation coefficient ($\rho$) and Regression Coefficient (RC) suggest stronger correlation and agreement with the reference standards. Across datasets, the model demonstrated consistently high performance, with $\rho$ exceeding 0.94 for LA and 0.98 for RA in the Utah dataset, indicating strong linear correlation. The lowest errors and biases were generally observed in Utah, while the Kobe dataset presented larger errors, particularly for RA volume, where MAPE reached 15.713\%. Despite this, the model still achieved reasonable agreement, with $\rho$ of 0.818 and RC of 0.666 in that case.

\subsubsection{Atrial Diameter Measurement}

Using segmentation outputs generated by the TASSNet framework, atrial diameter measurements were computed for both the LA and RA. Automated extraction from the segmentation masks can reduce manual measurement steps and may support reproducible large-scale analysis, subject to expert review and further clinical validation.

\begin{table}[!h]
    \centering
    \caption[Quantitative evaluation of atrial diameter measurements (in mm) on the Utah, Waikato, and Kobe datasets]{Quantitative evaluation of atrial diameter measurements (in mm) on the Utah, Waikato, and Kobe datasets for both the left atrium (LA) and right atrium (RA). Metrics include Mean Absolute Error (MAE), Root Mean Squared Error (RMSE), Mean Absolute Percentage Error (MAPE), Pearson correlation coefficient ($\rho$), Regression Coefficient (RC), and bias.}
    \label{tab:diameter}
    \resizebox{0.6\columnwidth}{!}{
    \begin{tabular}{lcccccc}
       \toprule
         Metric & \multicolumn{2}{c}{Utah} & \multicolumn{2}{c}{Waikato} & \multicolumn{2}{c}{Kobe} \\
         \cmidrule{2-7}
         & LA & RA & LA & RA & LA & RA \\
         \midrule
         MAE & 3.855 & 2.675 & 6.289 & 3.853 & 8.073 & 5.623 \\
         RMSE & 4.819 & 3.344 & 7.861 & 4.816 & 10.091 & 7.029 \\
         MAPE (\%) & 4.391 & 3.260 & 6.114 & 5.206 & 8.163 & 6.356 \\
         $\rho$ & 0.868 & 0.968 & 0.849 & 0.952 & 0.896 & 0.753 \\
         RC & 0.619 & 0.884 & 0.456 & 0.826 & 1.500 & 0.484 \\
         Bias & -3.252 & -1.543 & -1.829 & 2.436 & -1.141 & -3.481 \\
         \bottomrule
    \end{tabular}}
\end{table}

For diameter estimation, all values are in millimetres (mm). Similar to volume estimation, lower MAE, RMSE, MAPE, and bias indicate better accuracy, while higher $\rho$ and RC values suggest stronger correlation. The Utah dataset achieved the smallest errors, with RA $\rho$ reaching 0.968. The Kobe dataset showed larger errors, particularly for LA diameter (MAPE 8.163\%), though $\rho$ values still reflected a reasonable degree of correlation.

\section{Discussion}

{\subsection{Performance of TASSNet and Comparison with Baseline Models}}

{
TASSNet with ResNeXt encoders achieved competitive performance across the Utah, Waikato, and Kobe datasets and obtained the highest Dice values for many of the evaluated structures. However, the cross-centre results were structure dependent rather than uniformly robust. Cavity segmentation transferred more consistently from Utah to Waikato and Kobe, whereas wall segmentation showed larger reductions and wider uncertainty, particularly in the Kobe cohort.

The comparison with nnU-Net variants, SegResNet, Swin UNETR, U-Mamba, and other baselines indicates that architectural family alone does not determine cross-domain behaviour. Dimensionality, encoder diversity, and dataset characteristics interact to influence performance. The 2D/3D ensemble and ResNeXt encoder provided the most stable overall Dice performance, but neither removed the sensitivity of thin atrial walls to acquisition differences.

These findings support the use of controlled multicentre benchmarking rather than relying only on single-domain accuracy. A conventional U-Net-based framework with carefully controlled training can be competitive with more complex architectures, while the external results reveal where apparent in-domain gains do not transfer. Statistical measures of inter-dataset divergence, including MMD and KS, were consistent with the larger performance reduction observed for Kobe. The results should therefore be interpreted as evidence of relative architectural behaviour under the evaluated domain shifts, not as proof of universal multicentre generalisation or clinical readiness.
}

\subsection{2D vs 3D vs Ensemble performance}

The comparative analysis of 2D, 3D, and ensemble configurations reveals nuanced performance patterns that vary across datasets and anatomical structures, highlighting trade-offs between spatial modelling capacity and generalisability.

In the Utah dataset, 3D models, particularly TASSNet with ResNeXt encoders, performed best on complex structures such as the LA wall. This reflects the strength of volumetric architectures in modelling continuity and subtle morphological variation. For the RA wall and both atrial cavities, however, 2D and 3D models performed similarly, indicating that larger and more homogeneous regions gain little from full 3D context.

In contrast, the Waikato dataset showed a clear advantage for 2D models in wall segmentation. This shift likely stems from scanner and protocol differences, which appear to affect models that depend on stable volumetric patterns. The simpler inductive bias of 2D architectures makes them more resilient when image appearance varies. As with Utah, both approaches produced comparable results for cavity segmentation.

The Kobe dataset further emphasised the strength of 2D processing, with 2D models outperforming 3D across most structures and architectures. The high slice count in this cohort enables 2D networks to benefit from rich in-plane detail and consistent through-plane context without the computational compromises inherent to 3D. A small exception was the LA cavity, where the 3D TASSNet–ResNeXt variant showed a slight edge, likely due to the stability of cavity geometry that benefits from volumetric context.

{
Across the three datasets, ensemble models generally provided the strongest or most stable overall performance, although the gains were not uniform across every structure or architecture. This pattern reflects the complementary nature of the two paradigms. The 3D models capture volumetric continuity, while the 2D models preserve detailed in-plane information and may be less affected by through-plane resolution differences. Their fusion can reduce prediction variance, but it does not consistently resolve local wall-boundary errors.
}

{
The different levels of improvement seen in Waikato and Kobe underscore this interaction. The ensemble produced only modest gains in Waikato, which was more closely aligned with the training domain. Kobe introduced greater scanner and acquisition variability, and the ensemble improved several metrics by combining the different error patterns of the 2D and 3D models. Nevertheless, substantial wall-segmentation errors remained.
}

{
In this study, several 3D configurations were more sensitive to scanner-specific geometry and anisotropic spacing, while 2D models transferred more consistently in some external comparisons but lacked explicit cross-slice context. Ensemble fusion partially balanced these behaviours by combining heterogeneous inductive biases. The ResNeXt backbone was also associated with more stable overall performance than ResNet in the evaluated experiments. These observations are specific to the studied cohorts and require confirmation in larger external datasets.
}

{
Collectively, these comparisons indicate that dimensionality should be selected with the evaluation domain and image resolution in mind. The 3D configurations offered strong in-domain performance, whereas 2D or hybrid configurations transferred more consistently in several external comparisons. Combining heterogeneous inductive biases may therefore be useful for future cross-centre studies, but it should not be treated as a substitute for harmonisation, domain adaptation, or prospective validation.
}

\subsection{Failure Modes and Boundary-Level Challenges}

Despite the overall strong performance of TASSNet and its ensemble configurations, atrial wall segmentation remains a persistent challenge, especially when compared to cavity segmentation. This limitation stems from the intrinsic anatomical complexity of the atrial walls, which are thin, irregularly shaped, and often exhibit low contrast in LGE-MRI. These properties increase susceptibility to partial volume effects and make boundary delineation particularly difficult, especially in datasets with lower through-plane resolution or scanner-induced variability.

This difficulty is further reflected in the degradation of boundary-specific metrics such as ASD and the HD95. While the Dice score provides a measure of global overlap, it can obscure small but clinically significant boundary errors. In contrast, ASD and HD95 are more sensitive to local inaccuracies, particularly in regions with irregular contours or thin structures like the LA wall. These errors can directly impact downstream applications such as atrial wall thickness estimation, fibrosis quantification, and ablation therapy planning, where boundary precision is crucial.

Although ensemble models typically offer robustness by integrating the strengths of both 2D and 3D networks, their effectiveness is not universal. For instance, the lack of improvement or even slight degradation observed in some configurations, such as the SegResNet ensemble on the Waikato dataset, underscores the importance of architectural compatibility and dataset-specific behaviour. In some cases, ensemble fusion may introduce conflicting predictions, particularly if the component models struggle with boundary localisation in different ways.

{The lower wall Dice in Kobe dataset can be attributed to the thin-wall, multi-class nature of the task. Because the atrial wall occupies only a small number of voxels, local boundary deviations or small discontinuities can substantially reduce Dice, even when the overall atrial structure is correctly localised. The qualitative failure cases indicate that errors are typically concentrated around anatomically ambiguous regions, including pulmonary veins, valve planes, and atrial openings, rather than representing complete wall segmentation failure. Nevertheless, robust wall segmentation under strong domain shift remains an unresolved challenge.}

{\subsection{Cross-Dataset Evaluation Without Fine-Tuning}}

{
The cross-dataset evaluation characterises how a model trained on Utah behaves on Waikato and Kobe without target-domain fine-tuning. Cavity segmentation retained comparatively high Dice values across the external cohorts, whereas wall segmentation showed larger reductions and wider confidence intervals. This distinction is important because transfer performance cannot be described by a single overall measure when the anatomical targets differ substantially in size and boundary visibility.

The wall-performance reduction is consistent with the effects of low contrast, partial volume, anatomical variability, and differences in scanner hardware and acquisition parameters. The external results therefore indicate partial transfer for cavity structures but unresolved sensitivity for thin atrial walls. Larger external cohorts and prospective evaluation are required before drawing conclusions about general clinical use.
}

{
\subsection{Clinical Interpretation and Potential Future Impact}

Automated bi-atrial segmentation has potential value as a prerequisite for structural measurements and downstream analyses in AF. In the present study, cavity segmentation transferred more consistently across centres than atrial wall segmentation and supported retrospective estimation of atrial volume and diameter. These findings suggest that cavity predictions may provide useful initial segmentations for expert-reviewed structural analysis. However, prospective clinical usability was not evaluated.

Separate delineation of the LA and RA walls could support future chamber-specific analyses of wall morphology and fibrosis distribution. Nevertheless, fibrosis burden, ablation-target identification, recurrence prediction, and clinical outcomes were not assessed. The reported anatomical segmentation performance should therefore not be interpreted as evidence of improved fibrosis quantification or treatment planning.

The external results also suggest that expected correction requirements would differ by structure and imaging domain. Cavity predictions may require less correction in many cases, whereas wall predictions are more likely to require case-specific review, particularly near low-contrast boundaries, pulmonary veins, atrial appendages, valve planes, and in strongly shifted cohorts such as Kobe. Expert editing time and the extent of required correction were not measured, so no conclusion can currently be made regarding workload reduction or operator performance.

Accordingly, TASSNet should presently be interpreted as a benchmarking and assisted-segmentation framework rather than a clinically validated automated system. Future studies should quantify expert correction time, correction location and magnitude, inter-observer variability, fibrosis-burden error, and the effect of segmentation quality on ablation-planning and outcome-prediction tasks.
}

\subsection{Limitations and Future Directions}

The findings underscore the need for architectures and training strategies that better reflect the anatomical complexity of the atria. Components such as ResNeXt encoders, a two-stage pipeline, Dice Focal Loss, and Instance Normalisation collectively improved model stability, yet several structural challenges remain unresolved.

A consistent weakness across configurations was the segmentation of the atrial wall. Its thin geometry, low contrast, and irregular boundaries produced higher boundary errors, indicating that future architectures should incorporate mechanisms tailored to fine-scale structure. Potential directions include multi-scale fusion, attention modules, deformable convolutions, or boundary-aware refinement blocks to improve interface delineation. Incorporating anatomical priors or lightweight post-processing may further stabilise predictions at critical tissue borders. { Moreover, the discrepancy between Dice and surface-distance metrics indicates that overlap-based optimisation alone may be insufficient for thin-wall atrial segmentation. Future work should therefore explore boundary-aware losses, surface-distance objectives, anatomical priors, uncertainty-guided correction, and domain harmonisation strategies to reduce local boundary errors and HD95 outliers.}

{
Transferability also remains a key issue. Models performed well in domains aligned with the training data, but performance dropped on datasets with distinct acquisition characteristics. Although the present study did not apply intensity harmonisation or domain adaptation, the observed behaviour highlights an opportunity to address these shifts more directly. Future research should investigate image harmonisation, domain-aware augmentation, domain adaptation, test-time adaptation, uncertainty-guided correction, and the integration of weakly labelled or unlabelled target-site scans, particularly for wall segmentation in strongly shifted cohorts such as Kobe.
}

{
The analysis was limited to deep learning frameworks and three LGE-MRI datasets. Although the datasets represent multiple institutions, their overall protocol diversity remains restricted. The external cohorts were small, so their results should be interpreted as preliminary evidence of cross-domain behaviour rather than definitive proof of multicentre robustness. The bootstrap confidence intervals quantify this uncertainty and are wider for several wall and surface-distance metrics. Larger and more heterogeneous external cohorts will be required before drawing strong conclusions about clinical generalisability. Future work should also examine relationships between statistical divergence, predictive uncertainty, and segmentation behaviour.
}

{
Clinical evaluation in this study was limited to retrospective atrial volume and diameter measurements. Fibrosis-burden estimation, ablation planning, recurrence prediction, clinical outcomes, and expert editing time were not evaluated. In addition, only one final consensus segmentation was retained for each scan. Independent repeated annotations were unavailable, so inter-observer and intra-observer variability could not be quantified. These limitations should be addressed before the framework is considered for clinical use or claims are made regarding reduced correction workload or improved fibrosis quantification.
}

Overall, the results highlight a clear direction for advancement: improving boundary sensitivity, enhancing cross-domain stability, and expanding the methodological scope to ensure reliable performance across varied clinical environments.

\subsection{Toward Robust and Generalisable Atrial Segmentation}

This study reframes bi-atrial segmentation as a problem of generalisation rather than isolated model optimisation. By benchmarking multiple architectures under consistent conditions and across independent datasets, we demonstrate that apparent gains in segmentation accuracy often do not translate to robustness under domain shift.

{
Four key insights emerge from this analysis:
\begin{itemize}
    \item The 3D architectures provided strong spatial continuity but were more sensitive to scanner and resolution differences in several external comparisons. The hybrid 2D/3D ensemble mitigated some, but not all, of these effects.
    \item The ResNeXt encoder was associated with more stable overall performance than ResNet in the evaluated experiments, although thin-wall delineation remained challenging.
    \item Dataset shifts were accompanied by measurable degradation in wall segmentation, reinforcing the need for multicentre validation before clinical interpretation or deployment.
    \item The framework can serve as a baseline for future studies of harmonisation, transfer learning, domain adaptation, and uncertainty-aware correction. The benchmarking pipeline is architecture agnostic and can be extended to other model families.
\end{itemize}

In summary, this work provides a reproducible benchmark and a conceptual foundation for evaluating cross-domain segmentation in cardiac MRI. By emphasising systematic analysis and explicit limitations, it supports future development toward clinically reliable and transparent atrial-imaging tools.
}

\section{Conclusion}

{
This study presented TASSNet as a two-stage framework and benchmarking platform for bi-atrial wall and cavity segmentation from LGE-MRI. By evaluating 2D, 3D, and ensemble configurations with ResNet and ResNeXt encoders across Utah, Waikato, and Kobe datasets, the study quantified how architectural dimensionality, encoder design, and dataset shift influence segmentation performance.

The results show that atrial cavity segmentation generalises more reliably across centres than atrial wall segmentation. The 2D/3D ensemble and ResNeXt encoder provided the most stable overall Dice performance, but wall segmentation remained sensitive to domain shift, particularly in the Kobe cohort. These findings indicate that the framework is useful as a reproducible benchmark and as a starting point for assisted segmentation workflows. Assessment and validation as a clinical tool is part of ongoing research.

Future work will focus on larger external validation cohorts, boundary-aware segmentation, uncertainty-guided correction, domain adaptation or harmonisation, and direct validation against fibrosis burden, ablation planning measures, clinical outcomes, and expert editing time.}

\clearpage % Start on a new page
% \section*{CRediT authorship contribution statement}
% \printcredits

\section*{Ethics Approval and Consent to Participate}
Separate institutional ethical approval was received for each dataset.

 \bibliographystyle{elsarticle-num} 
 \bibliography{sn-bibliography}
% \begin{thebibliography}{00}

% %% For numbered reference style
% %% \bibitem{label}
% %% Text of bibliographic item

% \bibitem{lamport94}
%   Leslie Lamport,
%   \textit{\LaTeX: a document preparation system},
%   Addison Wesley, Massachusetts,
%   2nd edition,
%   1994.

% \end{thebibliography}
\end{document}